\documentclass{sig-alternate-05-2015}

\usepackage{algorithm}
\usepackage{algorithmic}
\usepackage{caption}
\usepackage{url}
\usepackage{boxedminipage}
\usepackage{xcolor}
\usepackage{makecell}

\usepackage{graphicx}
\usepackage{balance}  
\usepackage{amsmath}
\usepackage{amssymb}
\usepackage{listings}
\usepackage{graphicx}
\usepackage{paralist}
\usepackage{enumerate}
\usepackage{xspace}
\usepackage{subcaption}
\usepackage{caption}
\usepackage{comment}
\usepackage{array}
\usepackage{booktabs}
\usepackage{times}
\usepackage{inconsolata}
\usepackage{float}
\usepackage{tcolorbox}
\usepackage{placeins}
\usepackage[disabled]{aplcomments}

\usepackage{xcolor}
\usepackage{cmap}
\usepackage[T1]{fontenc}
\newcommand{\figlabel}[1]{\label{f:#1}}
\newcommand{\seclabel}[1]{\label{s:#1}}
\newcommand{\lstlabel}[1]{\label{lst:#1}}

\newcommand{\secref}[1]{Section~\ref{s:#1}}  
\newcommand{\figref}[1]{Figure~\ref{f:#1}}     
\newcommand{\listref}[1]{Listing~\ref{#1}}
\newcommand{\tableref}[1]{Table~\ref{#1}}

\newcommand{\lstref}[1]{Line~\ref{lst:#1}}

\usepackage{color,calc}
\definecolor{shade}{gray}{0.9}
\newsavebox{\tempboxa}
\newenvironment{mybox}
    {\begin{center}
		\vspace{-0.1in}
		\begin{lrbox}{\tempboxa}%
		\begin{boxedminipage}[h!]{\columnwidth}
		}
		{
		\end{boxedminipage}
		\end{lrbox}%
    \fcolorbox{black}{shade}{\usebox{\tempboxa}}
		\vspace{-0.1in}
    \end{center}
    }

\newcommand{\Action}{controller action\xspace}
\newcommand{\Actions}{controller actions\xspace}
\newcommand{\orms}{ORM frameworks\xspace}
\newcommand{\orm}{ORM framework\xspace}
\newcommand{\tool}{\textsc{CADO}\xspace}

\newcommenter{cong}{0.4,1.0,1.0}
\newcommenter{alvin}{1.0,1.0,0.4}
\newcommenter{shan}{0.8,0.1,0.8}

\begin{document}






%
\title{Database-Backed Web Applications in the Wild:\\How Well Do They Work?}

%
%
%
%
%

\author{
\end{tabular} 
\def\{{\char123}
\def\}{\char125}
\begin{tabular}{c@{\qquad}c}
\aufnt
Cong Yan & Shan Lu\\[0.5ex]
Alvin Cheung & \\[1ex]
\affaddr University of Washington &
\affaddr University of Chicago 
\end{tabular}
\begin{tabular}{c} 
}
\numberofauthors{2}


\maketitle

\begin{abstract}
Most modern database-backed web applications are built upon
Object Relational Mapping (ORM) frameworks.
While \orms ease application development by abstracting persistent data
as objects, such convenience often comes with a performance cost.
%
In this paper, 
we present \tool, a tool that analyzes the application logic and its interaction
with databases using the Ruby on Rails \orm. 
\tool includes a static program analyzer, a profiler and
a synthetic data generator
to extract and
understand application's performance characteristics.
We used \tool to 
analyze the performance problems of  
27 real-world open-source Rails applications, covering domains such as online forums, 
e-commerce, project management, blogs, etc.
Based on the results, we uncovered a number of 
issues that lead to sub-optimal application performance, ranging
from issuing queries, how result sets are used, and 
physical design. 
We suggest possible remedies for each issue, and highlight 
new research opportunities that arise from them.
\end{abstract}

\section{Introduction}
Object-relational mapping (ORM) frameworks are widely used 
to construct applications that interact with database
management systems (DBMSs). Despite the many implementations
of such frameworks (e.g., Ruby on Rails~\cite{rubyonrails}, Django~\cite{django},
Entity Framework~\cite{entityframework}, Hibernate~\cite{hibernate}), 
the design principles and goals 
remain the same: rather than embedding SQL queries into application
code, 
\orms present persistently stored data as 
heap objects and let developers manipulate persistent data
in a similar way as regular heap objects via ORM APIs~\cite{ejb, 
activerecord}. These API calls are translated by the \orms into
queries and sent to the DBMSs. By raising the level of abstraction, this approach 
allows developers to implement their entire application using
a single programming language, thereby greatly enhancing code 
readability and maintainability.

However, the increase in programming productivity comes at a cost.
By hiding DBMS details behind an API, \orms lack accesses to 
high-level semantic information of the application, such as how persistent
data is used. In addition, abstracting queries
into API calls prevents application
compilers from optimizing these calls and related computation, 
as compilers treat these API calls
as unknown external functions that cannot be altered and
conduct arbitrary modifications to related objects

\begin{figure}[h]
  \centering
  \includegraphics[page=4, width=1\columnwidth]{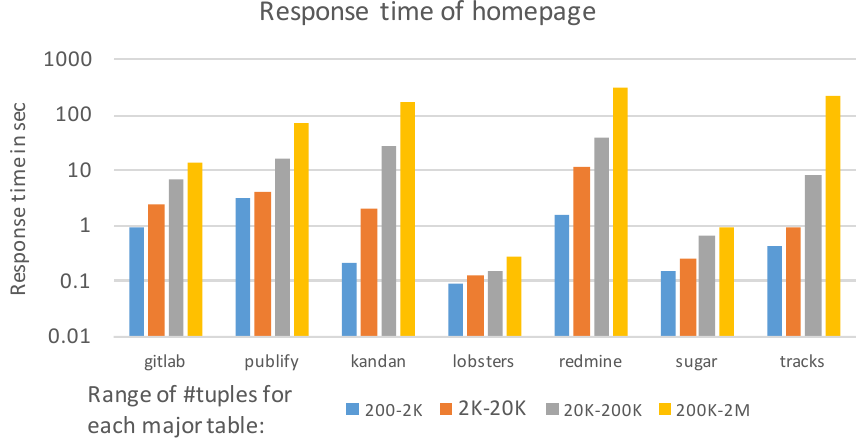}
  \caption{How response time of homepage ({\bf log-scale} in Y-axis)
scales with database size (ranging from 1M to 25G).}
  \figlabel{homepage}
\vspace{-0.1in}
\end{figure}

Both aspects make applications built
atop \orms vulnerable to performance problems.~\figref{homepage} shows that
for many applications, the response time for some applications
becomes intolerable (more than 10 sec) 
when the number of records is beyond 200K. 
As application performance 
is often of critical concern for many ORM applications 
and the data processed by applications is growing,
the performance problems make these applications
hard to handle nowadays big-data challenges.
\alvin{Why bring this issue up at this point? It sounds like a scaling issue to me.}
\cong{I just want to intuitively show the performance problem is serious, and make the application
really slow. The problems doesn't usually show up when the data is small.}
\alvin{also, I thought this paper is not entirely about performance? 
For instance consider the testing section.}

Some problems are well-known in the data management
~\cite{speedup-activerecord},  
software engineering research communities
~\cite{chen:se14:antipattern, chen:finding}, and 
developer communities as well~\cite{Nplus1}. 
Lacking automatic tools to solve these problems, 
developers often need to ``hand-tune'' their ORM-based applications,
for instance, implement caches for persistent 
data~\cite{orm-caching, rails-caching} or customize 
physical design for each application~\cite{orm-mapping, inheritance-mapping}. 
Doing so greatly complicates the application and undermines the 
original purpose of using~\orms.

Addressing performance issues associated with ORM-based applications
requires an understanding of how such applications are built and used, which
in turn requires examining the application, the \orm, and the DBMS in 
tandem. While there has been prior work in optimizing ORM-based 
applications~\cite{chen:se14:antipattern, chavan:db11:dbridge, chen:finding},
they focus on specific performance issues,
and evaluated on a small range of applications.
It is unclear how widely spread are such issues are among real-world
ORM-based applications, or whether other issues exist.
Various tools for detecting and measuring 
application performance target either the application or the 
DBMS, and we are unaware of any that examines both.

In this paper, we describe \tool, a new tool that analyzes ORM-based
applications by examining the application, the \orm, and 
the DBMS that the \orm interacts with. 
\tool currently works on ORM-based applications that are built using 
the Ruby on Rails (Rails) framework, due to its popularity among 
developers.
\tool has two components:
\begin{compactenum}
\item A static code analyzer that not only performs traditional program analysis
but also builds a new data structure {\em Action Flow Graph} (AFG),
which captures the control flow, data flow and interactions
among DBMS, application server and user.
\item A dynamic profiler that executes the application 
and collects runtime information such as the state of the database 
and execution time.
\item A data generator
based on users' interactions with the application to populate database. 
\end{compactenum}

We use \tool to perform, to the best of our knowledge, the first comprehensive
study of ORM-based applications ``in the wild''. Specifically, we collect 27
 real-world applications that are built using Rails and 
study them with \tool. We purposefully chose applications
based on their popularity and domain. 
The results show that many performance issues still remain:
many queries retrieve much redundant data; query results are unnecessarily returned
to the application; many queries repetitively run the same computation;
many queries return large number of records which make later data processing
slow; much effort is spent on translating relational data to other 
data structures, etc.
For each observation, we point out the direction for optimizations and
show potential performance improvement by examples.

In summary, this paper makes the following contributions:
\begin{itemize}
\item We built \tool, a tool for analyzing applications 
built with Ruby on Rails ORM framework. \tool studies
applications using static analysis and generates 
action flow graphs, which describe the flow of 
persistent data and user interactive data among
web browser, application server, and database systems.
It also performs dynamic analysis and profiles applications using 
synthetically generated data.
\item We used \tool to perform the first comprehensive study of 
27 real-world applications covering a wide variety of domains including forums,
social networking, e-commerce, etc. 
Our results show that many applications expose similar patterns,
in aspects including issuing queries, how query results are used,
scalability with database size, caching and prefetching, and physical design.
Most of our findings have not been studied or only partially
studied by previous work.
\item Based on our observations, we propose possible solutions to 
the performance problems and optimizations,
which reveals many new research
opportunities.  
\end{itemize}

In the following 
we review the design of \orms and how they are used to 
construct database-backed applications
in~\secref{background}.  In~\secref{methodology}, we
describe \tool and our study. Then in~\secref{queryopt},
\secref{physical},~\secref{scalable},~\secref{caching_and_prefetching}
and~\secref{testing},
we present the findings from our study and suggest optimizations. 

\section{backgroud}
\seclabel{background}

In this section we 
give an overview of Ruby on Rails (we refer to as ``Rails'' below), 
the architecture of applications built using Rails,
and how Rails implements object relational mapping.

\begin{figure}[h]
	\centering
	\includegraphics[page=3, width=0.8\columnwidth]{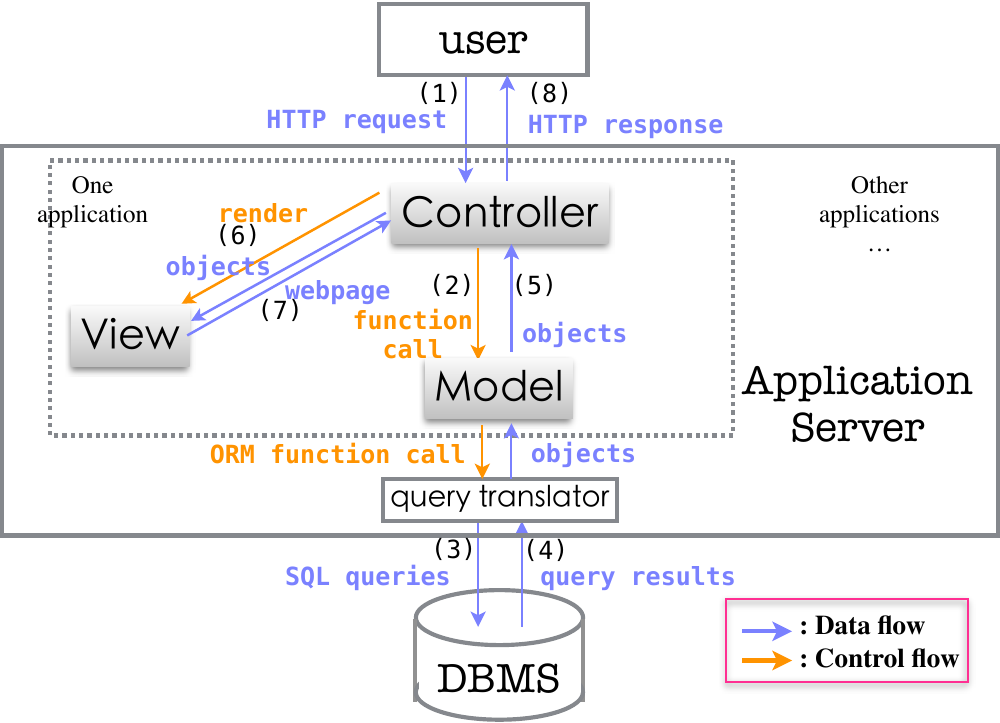}
	\caption{Architecture of a Rails application}
	\figlabel{ormarchitecture}
\vspace{-0.1in}
\end{figure}

\begin{figure*}
	\centering
	\includegraphics[page=2, width=1.95\columnwidth]{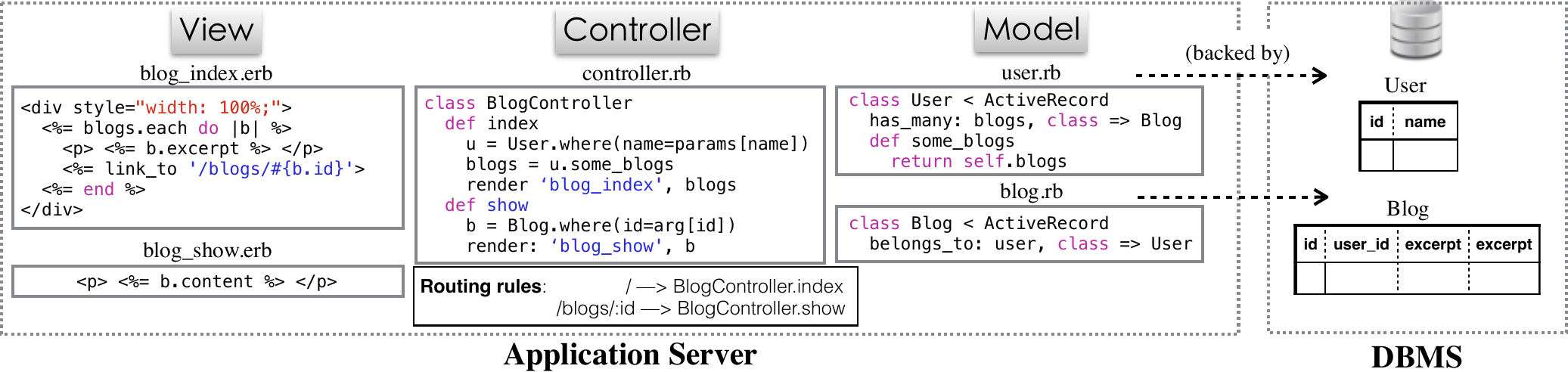}
	\caption{Rails application example}
	\figlabel{mvcCode}
\vspace{-0.1in}
\end{figure*}

\subsection{Design of Rails applications}
\seclabel{rails_app}
We chose Rails applications as representatives of database applications,
given the popularity of the Rails framework and the number of 
open-source and commercial applications (e.g., Airbnb, Shopify, Hulu, etc)
that are constructed using it.
\cong{Need to mention somewhere how our findings applied to other ORMs...}
\figref{ormarchitecture} shows the typical structure of 
a typical application built with Rails, with the application hosted
on the Rails application server that communicates with the DBMS
as needed when it executes. This two-tier architecture is common 
across applications that are constructed using other \orms, such as
Hibernate~\cite{hibernate}, EntityFramework~\cite{entityframework}, and 
Django~\cite{django}. 

Internally, a Rails application is organized into 
model, view, and controller components.
Upon receiving a user's request, say to render a web page, 
(step 1 in~\figref{ormarchitecture}), the Rails server invoke the
corresponding
{\it action} that resides in the appropriate 
application {\it controller} based on the 
routing rules provided by the application (shown in~\figref{mvcCode}).
During its execution, the controller interacts with the DBMS by
invoking Rails-provided ORM functions (2), which Rails translates into
SQL queries (3).
%
Upon receiving query results (4), data is serialized into 
{\it model} objects and that are 
returned to the controller (5). 
The retrieved objects are then passed
to the {\it view} (6), which constructs the webpage (7) that is returned 
to the user in response (8).

As an illustration, \figref{mvcCode} shows the code of a
blogging application.
While the controller and model are written in Ruby, 
the view code is implemented using 
a mix of ruby and mark-up languages such as HTML or Haml. 
There are two actions defined in the controller: {\tt index}
and {\tt show}. Those actions call the ORM functions {\tt
User.where}, {\tt User.some\_blogs}, and {\tt Blog.where} that are translated
to SQL queries for data retrieval. 
The action {\tt Blog.index} passes the
blogs retrieved from the DBMS 
to {\tt render} to construct the webpage defined by 
the view file {\tt blog\_index.erb},
where the excerpts ({\tt b.excerpt}) of all blogs and their 
contents ({\tt b.content}) are rendered 
as part of the generated webpage. 
For each blog, a url link to the page showing  
that blog ({\tt link\_to}) is also listed on the webpage.
Clicking on the link will bring the user to a separate page, with 
another action ({\tt Blog.show}) triggered. 
\alvin{figure 2 has inconsistent tab spaces. Some are 4 and some are 2}

\subsection{Object relational mapping in Rails}
\seclabel{tableMapping}

Like many other \orms, Rails by default 
maps each model class directly derived from ActiveRecord~\cite{activerecord}
into a
single table. 
As shown in~\figref{mvcCode}, {\tt User} objects are stored in the 
{\tt User} table in the DBMS, and likewise for {\tt Blog} objects.
Programmers have the option of specifying which subset of fields in a model class 
should be physically stored in the corresponding database table, and
defining relationships among model classes via association,
such as {\tt belongs\_to}, 
{\tt has\_one} and {\tt has\_many}~\cite{rails:association}.
For example, in~\figref{mvcCode}, 
each {\tt Blog} object {\tt belongs\_to} one 
{\tt User}, hence Rails stores the unique 
{\tt user\_id} associated with each {\tt Blog}.
\alvin{what is the point of bringing association up?}
\cong{In sec 4.4 we talk about loading associated data...}

\section{Architecture of \tool}
\seclabel{architecture}

In this section, we describe the architecture of \tool 
and the application corpus that was chosen for our study.
As mentioned, \tool consists of a static analyzer, an application profiler
and a data generator.
. 
\alvin{I have never heard of 'dynamic profiler.' Does 'static profiler'
even make sense?}
In the following, we describe how each component works.

\subsection{Program analyzer}
\seclabel{staticAnalysis}

\begin{figure}[h]
	\centering
	\includegraphics[page=1, width=1\columnwidth]{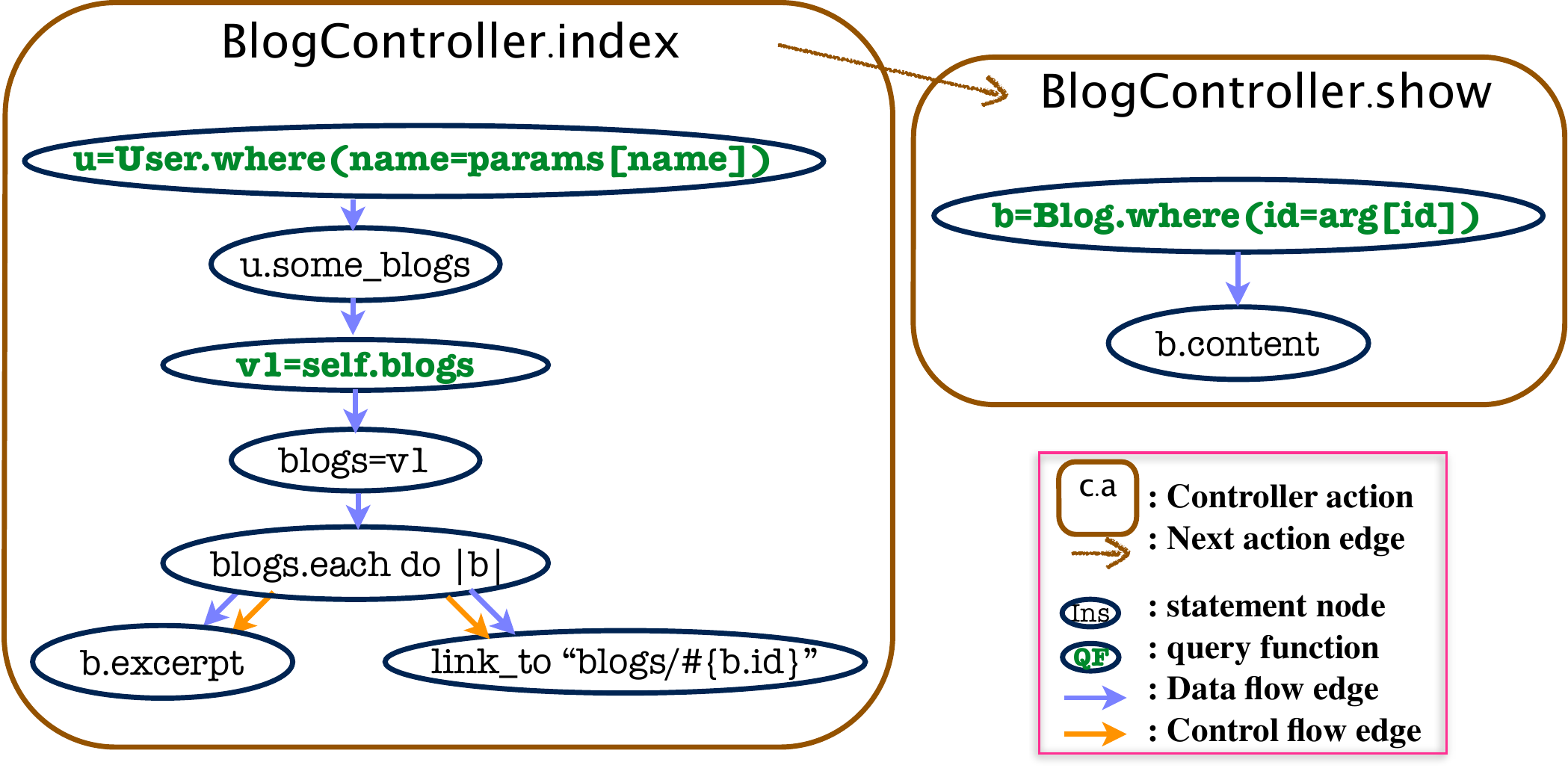}
	\caption{Action Flow Graph (AFG) example}
	\figlabel{afg}
\end{figure}

The static code analyzer in \tool takes in Rails application source code,
examines its interaction with the DBMS, and generates an {\it Action Flow Graph}
(AFG) for each application. 
AFGs are inspired by Program Dependence 
Graphs (PDG)~\cite{pdg}
that capture control and data flow in general-purpose programs.
In addition, AFGs contain {\it next action} edges 
between pairs of \Actions $(c_1,c_2)$, meaning that  
$c_2$ can be invoked from $c_1$ as a result of user 
interactions (e.g., by clicking on a link on the view generated
from action $c_1$).

To construct AFGs, 
\tool first builds the PDG for each \Action. \tool relies on
JRuby~\cite{jruby:instr} to construct control-flow graph, and
uses classical algorithms~\cite{dragonbook} to infer
data flow.
To perform inter-procedural analysis so that all functions
called by an action are considered,
\tool recursively inlines all function calls.\footnote{\small{\tool currently
assumes all recursive calls terminate with call depth 1.}}
In order to correctly inline, \tool performs type inference as Ruby
is dynamically typed~\cite{an:ace09:rubytype}.
To ensure all the code triggered by an actions are all inlined including
the code in view
\tool finds out the view file to be rendered for each {\tt render}
call in controller~\cite{activeview}.
Since view files are usually written in Haml or a mix language of ruby
and HTML, \tool converts non-ruby languages like Haml into ruby and drop
out HTML before inline.
\tool also inlines filter functions that are automatically
run before an action, and validations that are auto-run before 
executing any function that modifies database.

\tool then build edges to connect different actions.
To do so, \tool identifies functions that generate URL links 
(e.g., {\tt link\_to} in~\figref{mvcCode}), or generate a form
in the output webpage, and determine what action is triggered
by each of these links and forms by applying the application
routing rules (e.g., those listed in~\figref{mvcCode}).
Users can go to the next page or submit form
to trigger next action, so we construct a {\it next action} edge
between the action generating the URL/form, and the next action.



\subsection{Application profiler}
\tool comes with a profiler that runs
the application on a Rails server and instantiates 
a crawler as the client to profile the application.
The crawler starts by visiting
the homepage of the application, and
randomly chooses a link to visit next.
If the visited webpage contains a form, \tool 
fills in the form with random data in order to proceed to the next page. 
\tool currently only profiles each application using a single client crawler.
Meanwhile, \tool profiles the application running on the Rails server by 
recording the running time of each \Action as the crawler visits each of the 
pages. It also records the queries issued by the application to the DBMS, 
the time spent on each query, and the results returned by each query.

\subsection{Data generator}
\tool comes with a data generator for users to 
generate synthetic data and populate the DBMS that the 
given Rails application interacts with. 
To generate data, \tool visits webpages served by the application, 
and fills out forms embedded in the served pages with random data.
To ensure that data is populated appropriately, 
we 
manually examine the constraints of each field in the embedded
forms, for instance  
ensuring that only strings of the form
'xxxx-xx-xx' are generated when populating the `date' field of a form, 
since the application code parses such strings
and saves them as datetime to the database.
We tailor the data generator to fill the data
of correct format for such fields, and consider performing such
checks automatically to be future work.


\subsection{Limitations}

\tool currently has three
limitations. First, \tool's static analyzer only performs path insensitive
analysis, meaning that it considers both sides of a branch to be taken.
\tool also considers each loop to be executed once.
Second, \tool currently does not model user behavior. As a result, 
the static analysis performed by \tool might not reflect the frequency
of actions, and the generated data might not be representative of 
actual data 
when the application is deployed, even though all generated data is guaranteed
to satisfy the constraints imposed by the application.
In addition, \tool performs only a limited form of inter-action analysis. 
In particular, \tool currently does not perform multi-user and multi-session
analysis.

\section{Study parameters}
\seclabel{methodology}
\begin{table*}
\small
\centering
	\begin{tabular}{rr|rr|rr|rr|rr|rr|rr}
	\toprule
	App & Loc & App & Loc & App & Loc & App & Loc & App & Loc & App & Loc & App & Loc \\
	\midrule
	\multicolumn{2}{c|}{\textbf{forum}} & \multicolumn{2}{c|}{\textbf{social networking}} & \multicolumn{2}{c|}{\textbf{collaboration}} &\multicolumn{2}{c|}{\textbf{task management}} & \multicolumn{2}{c|}{\textbf{resource sharing}} & \multicolumn{2}{c|}{\textbf{e-Commerce}} & \multicolumn{2}{c}{\textbf{blogging}}\\
\hline
	forem	&5957	&kandan	&1694	&redmine	&27589	&kanban	&2027	&boxroom	&2614	&piggybak	&2325	&enki	&5275\\
	lobsters	&7127	&onebody	&32467	&rucksack	&8388	&fulcrum	&4663	&brevidy	&13672	&shoppe	&5904	&publify	&16269\\
	linuxfr	&11231	&commEng	&34486	&railscollab	&12743	&tracks	&23129	&wallgig	&11189	&amahi	&8412 & & \\		
	sugar	&11738	&diaspora	&47474	&jobsworth	&15419	&calagator& 1328& &	 &sharetribe	&67169	& &\\	
	 & & & &gitlab	&145351	& & & & & & & & \\							
	\bottomrule
	\end{tabular}
	\caption{Application category and lines of code (source code 
           of all applications are hosted on github). 
All applicaionts listed are studied with static analysis.
Among them, gitlab, kandan, lobsters, publify,
redmine, sugar, tracks are profiled.} 
	\label{applicationCat}
\end{table*}

We conducted a large scale case study of 27 open-source Rails applications.
Applications were selected from github based on their popularity (80\% of them
have more than 200 stars), number of contributors (88\% of them have more than
10 contributors), number of commits, and category.
The applications, their categories, 
and lines of code are shown in~\tableref{applicationCat}.
Overall, they cover a broad range of  
characteristics in terms of DBMS usage: transaction-heavy (e.g., e-commerce),
mostly read-only (e.g., social networking),
write-intensive (e.g., blogging, forum), 
and data that can be horizontally partitioned based on users
(e.g., blogging, task management),
or heavily-shared among users
(e.g., forum, collaboration). We believe that these represent all major 
categories of ORM-based applications.
For findings presented in following sections,
we take the average of all actions for each application (shown in figures),
and then average across all applications (statistic in findings),
unless specified separately.

We deployed 7 of the above Rails application server on AWS with
4 CPUs, each with 16GB of memory,
and run client crawlers on PC when doing profiling.

\alvin{I moved the 'assumptions' paragraphs to sec 3.4 as I don't think 
those are assumptions about our study but rather limitations of \tool.}

\section{Improving Query Execution}
\seclabel{queryopt}
In this section, we report on our findings from using \tool
to analyze the applications described in~\secref{methodology}.
We categorize our findings into four aspects:

\begin{asparaitem}
\item Many queries share subexpressions as other
queries. 
Caching and reuse intermediate result \alvin{what is an intermediate result} 
on the server can potentially reduce
query execution time for subsequently overlapping queries. 
%
\item In many cases, 
the result of queries are only used to issue subsequent queries.
Combining such queries can reduce 
the amount of data to be transferred from the DBMS to the server.
%
\item Much of the query results are not used in the application.
Rewriting queries to return only the data needed can substantially
improve application performance.
\item Many database columns are computed from constant values.
Program analysis can figure out the domain of such columns
\alvin{please revise this. This doesn't make sense.} 
\end{asparaitem}

\subsection{Queries sharing subexpression}
\seclabel{sec:partial_pred}

\noindent{\bf Motivation.}
Some queries share common subexpressions, which cause
repetitive computation.
To improve performance, such queries can be rewritten such that
previously retrieved results are cached and reused.
An example is shown in~\listref{queryIntermediate}.
First, the {\tt issues} variable is created on~\lstref{issueSelect}.
After that, the {\tt any?} function call on~\lstref{issueAny} issues a {\tt COUNT} query.
Then the {\tt issues} are later rendered with their status information 
(\lstref{issueInclude}), issuing a {\tt JOIN} query.
The {\tt JOIN} query (Q2 in~\figref{queryIntermediateFig}) shares
selection and and the first join predicates
with the {\tt COUNT} query (Q1 in~\figref{queryIntermediateFig}).
\alvin{I thought the selection and join predicates are exactly the same.}

When executed separately, queries Q1 and Q2 take 6.45 seconds to finish.
We simulated the effect of caching intermediate results by creating a view
to store the results from a manually-written query with shared predicates
(Q3 in~\figref{queryIntermediateFig}), and changed the queries 
(as shown in Q4 and Q5) to use view {\tt v}. 
When using the cached results, the total query execution time is reduced
to 2.84 seconds (i.e., a 44\% reduction).

%

\noindent{\bf \tool Analysis.}
Since in Rails all queries are generated by ORM from object-oriented language,
we are able to analyze the program, instead of query log, to find out the shared 
query predicate. 
We used \tool to identify the ORM functions that are called to construct queries, 
and manually collected statistics regarding how frequent queries overlap.
For instance,
object {\tt issues} calls the Rails standard function {\tt any?}, 
which is translated to a {\tt COUNT} query.
If the object satisfies the following conditions: i) returned by query functions
that store subexpressions (e.g., {\tt Issue.include.where.group\_by}), and 
ii) calls
another query function (e.g., {\tt issues.includes} that translates to a {\tt JOIN} query),
then we count all the queries issued by that object (both {\tt issues.any?}
on~\lstref{issueSelect} and {\tt issues.includes} on~\lstref{issueInclude})
as "query sharing subexpressions".
\alvin{this is confusing because 1. you call an object a query (look at the 
subject of the above sentence); 2. which object is 'query' and which object is
'other query' in this case? The above sentence only refers to one object; and 
3. the example above mentions group by and join, and you call them sharing 
'partial predicates.' Where are the predicates?}
The average number of such queries in one action is shown in~\figref{queryPartial}.

\begin{lstlisting}[language=Ruby,caption={Example of queries partially share predicate.
 Code snippet abridged from redmine\cite{redmine}.},label={queryIntermediate}]
prj_pos='projects.lft > ? AND projects.rgt < ?'
issues=Issue.includes(:projects).where(prj_pos).group_by(:tracker_id) @\lstlabel{issueSelect}@
if issues.any? @\lstlabel{issueAny}@
	issues.includes(:statuses).each do |tid, i| @\lstlabel{issueInclude}@
    render_issue_group(tid, i)
\end{lstlisting}

\begin{figure}[h]
  \centering
  \includegraphics[page=2, width=1\columnwidth]{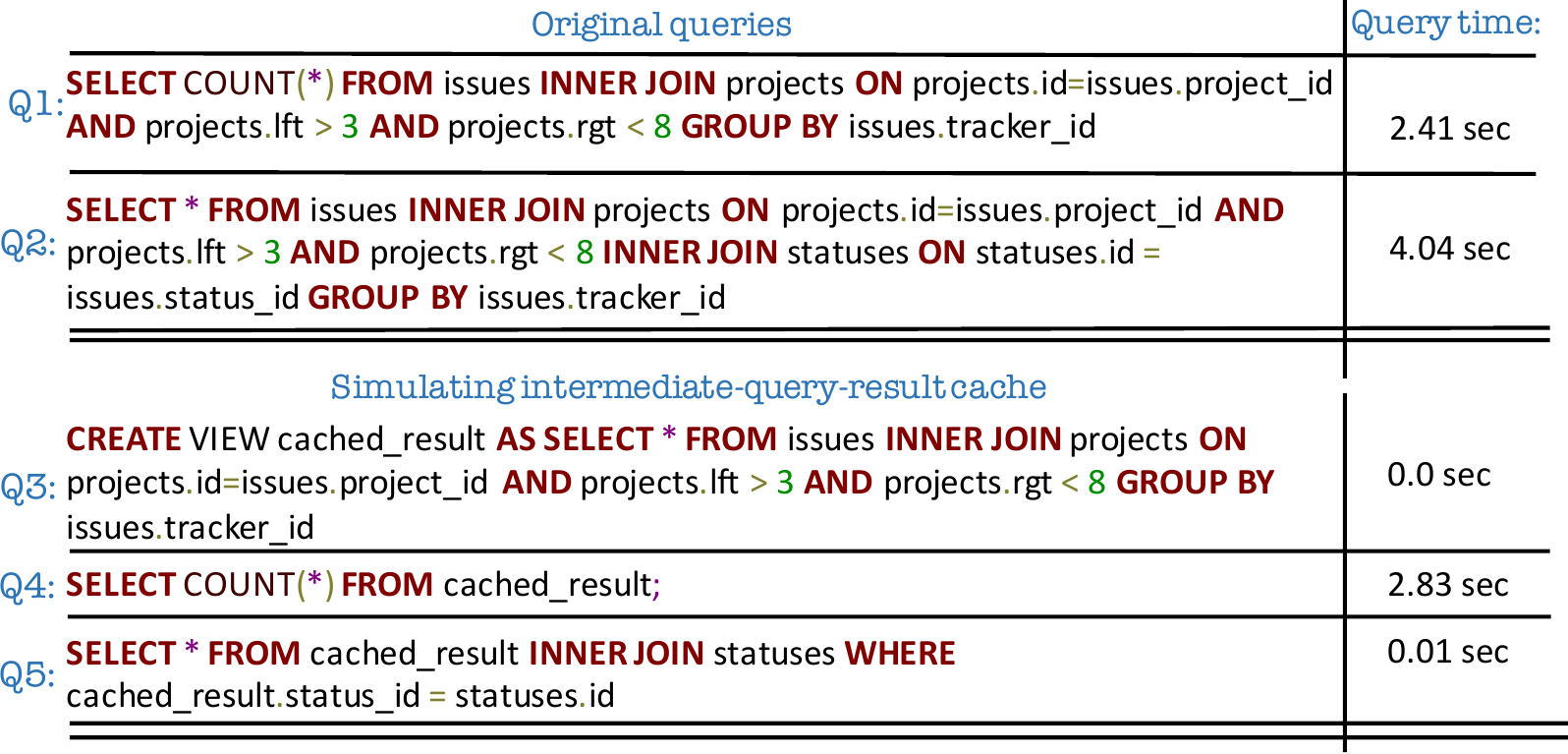}
  \caption{Performance gain by caching of query results. 
The issues/projects/statuses table has 100K/20K/10 records respectively.
}
  \figlabel{queryIntermediateFig}
\vspace{-0.1in}
\end{figure}

%

\begin{figure}[h]
  \centering
  \includegraphics[width=1\columnwidth]{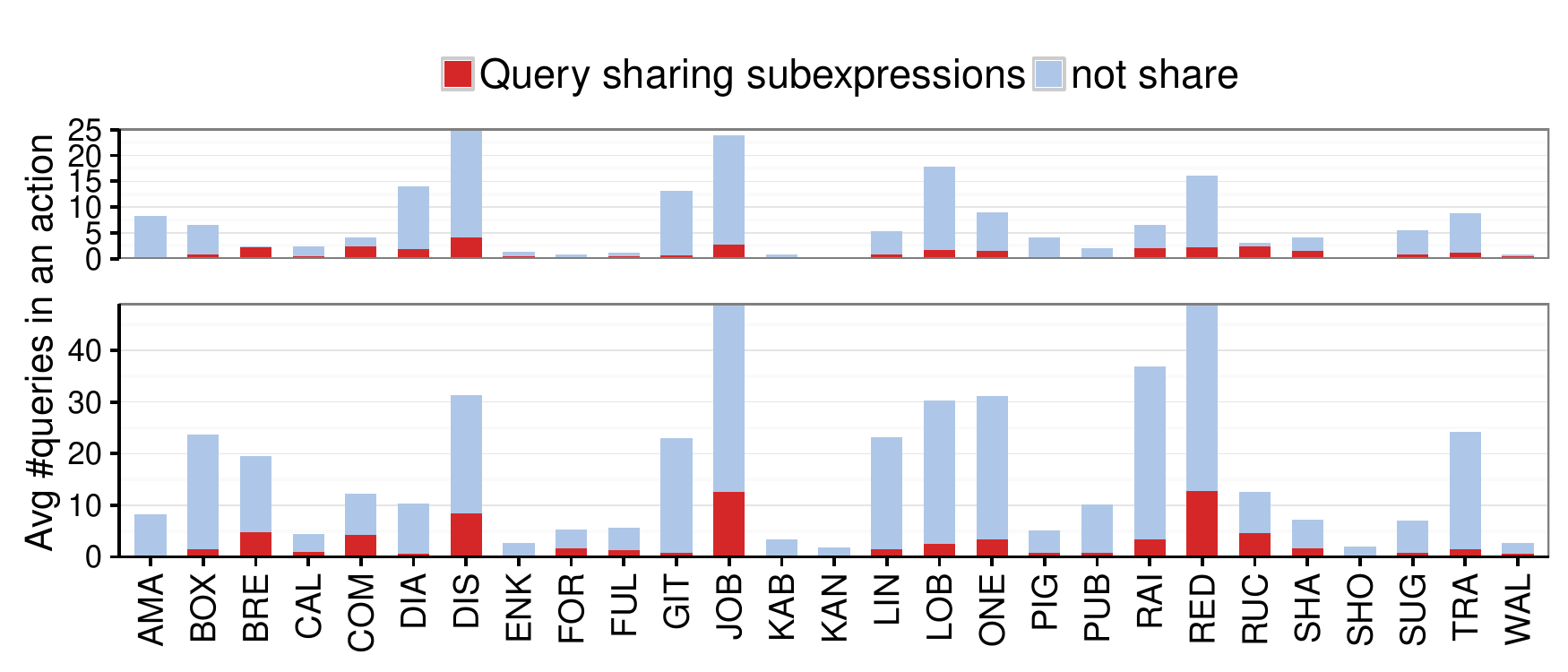}
  \caption{Queries partially share predicates with other queries. The upper figure shows the queries issued in loops, and the lower shows
queries not issued in loop. Similar for Figure 7, 13, 17. We separately
show these two types of queries since queries in loop may be repetitively
issued, potentially having greater impact on performance.}
  \figlabel{queryPartial}
\end{figure}

\noindent{\bf Findings \& Implications.}
From the application corpus, we observe that
16.9\% of the queries issued by the applications 
share subexpressions with other queries.
By manually checking these queries, we find that many
of these queries are performance critical ones
and are usually more complex than other queries.
Analyzing query logs or the code statically using similar analysis
as discussed above can identify queries such as that 
shown on~\lstref{issueSelect} in~\listref{queryIntermediate}. 
Separate views or tables can then be created such that subsequent
queries can leverage them to reduce query execution time 
(e.g., Q3 in~\figref{queryPartial}). However,
care must be taken to ensure that the cached results are updated or invalidated
appropriately. While this is dependent on user behavior that
\tool currently does not model, but from our manual code inspection
we believe that the cached data will be useful as many queries 
with shared subexpressions happen within the same user \Action
(i.e., the same transaction). \alvin{Check if this is true, I made that up.}

%
%

\noindent{\bf Prior work.}
Identifying shared subexpressions in the context of multi-query optimization
has been explored~\cite{commonexpression,qpipe}. 
Most prior work target analytical queries as they have
complex structures and are likely to have many common subexpressions. However, 
detecting subexpression is difficult using query logs due to mixing of queries from 
different clients and is costly as well. \cite{sharedb} presents
an approach to combine queries with shared computations to improve system throughput, 
but at the cost of slowing down individual queries. 

We propose an alternative way to detect common subexpressions by statically
analyzing the application code. 
Compared with analyzing query logs, 
this approach substantially reduces the effort in detection. 
We also show that this optimization
opportunity is common in ORM applications \alvin{16\% doesn't sound very common 
to me, unless these are performance critical ones?}, and techniques introduced in prior 
work on query caching~\cite{qpipe} can be applied.

\subsection{Unnecessary transfer of query results} 
\seclabel{unrequired_query}

\noindent{\bf Motivation.}
Transferring large query results 
from the DBMS to the application server
can cause significant performance degradation.
Using \tool, we observe that there is a significant number of queries whose
results are only used as parameters for subsequent queries and nowhere else.
As such, if such queries can be combined or grouped together as a stored
procedure, 
one way to reduce such data transfer and improve application performance when the 
query result returned is only used as parameters to some subsequent queries
is to group them into a stored procedure or combine them
as a single query to avoid transferring unnecessary results
from the DBMS.

\listref{issueSelect} shows such an example.
Query on~\lstref{select1} 
in the original implementation returns all members from group 1, the result of which 
({\tt m})
is only used for query on~\lstref{select2}. 
These two queries can be sent to the database as a stored procedure,
or combined into a single
query as shown in~\listref{issueJoin}. 
We tested the performance gain of query combining.
When {\tt Q1} returns 20K records (340KB) and network delay is 100ms, 
{\tt Q1} and {\tt Q2} takes 1.46sec and 1.3sec (2.76sec total) respectively.
When using stored procedure in which the same {\tt Q1} and {\tt Q2}
are executed, it takes 2.12sec in total, reducing 23.5\%; 
when issuing a join query, the query takes 1.02sec, reducing 62.3\%. 

\begin{lstlisting}[language=SQL, caption={Original queries listing issues by members from group 1. Code snippet abridged from redmine\cite{redmine}.}, label={issueSelect}]
Q1: m = SELECT * FROM members WHERE group_id = 1;  @\lstlabel{select1}@
Q2: SELECT * FROM issues WHERE creator_id IN (m.id) AND AND is_public = 1 @\lstlabel{select2}@
\end{lstlisting}

\begin{lstlisting}[language=SQL, caption={Combining Q1 and Q2 in Listing 2}, label={issueJoin}]
SELECT * FROM issues INNER JOIN members ON members.group_id = 1 AND issues.creator_id = members.id AND issues.is_public = 1 
\end{lstlisting}

%

\noindent{\bf \tool Analysis}
We target at queries whose results are only used to issue other queries
so that their results do not need to return to the application,
and we use static analysis to find out such queries.
To know how query result is used, we trace the dataflow from each query node
until we reach either a node with a query function, 
or a node having no outgoing dataflow edge. 
We refer to such nodes as the read query sinks. 
We subsequently categorize sinks into the following: 
(1) query parameters in subsequent queries; 
(2) rendered in the view; 
(3) used in a branch conditions; and 
(4) assigned to global variables.
After analyzing the sinks, 
we count the number of such queries whose result is only used for other queries
(category (1))
across all actions, 
and the result is shown in~\figref{queryOnlyToQuery}.

\begin{figure}[h]
  \centering
  \includegraphics[width=1\columnwidth]{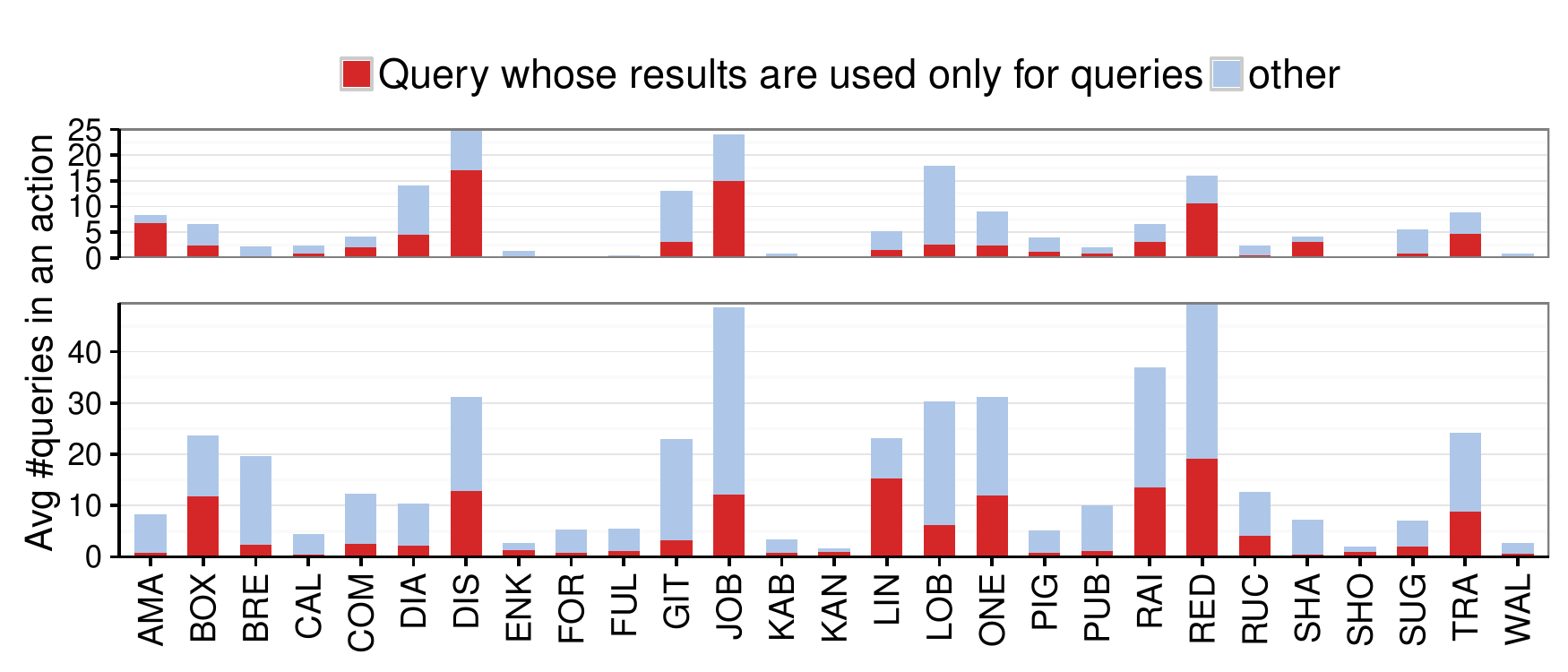}
  \caption{Queries whose result is only used in other queries.}
  \figlabel{queryOnlyToQuery}
\vspace{-0.1in}
\end{figure}

\noindent{\bf Findings \& Implications.}
We have observed that
35.4\% of queries return results that are only
used to issue other queries.

The above finding indicates that  
one can use static program analysis, similar to that performed 
by \tool,
to automatically identify queries that can be cominbed.
The compiler can use such information to automatically 
create stored procedures and avoid unnecessary data transfers.
Similar approaches have been explored earlier in prior work~\cite{pyxis}.
A query optimizer can also leverage such information to combine
multiple queries into a single one to further reduce query execution time.

\subsection{Unnecessary column retrieval}
\seclabel{sec:unne_column}

\noindent{\bf Motivation.}
Using \tool, we find that many queries issued by the applications retrieve
columns that are not used in subsequent computation. 
Clearly, automatically identifying
and avoiding unnecessary column retrieval can reduce both
query execution time and amount of data transferred.

\noindent{\bf \tool Analysis.}
\tool uses static analysis to identify the columns retrieved by each query,
and then identifies the columns that are used within the function where they 
are retrieved.
The number of unused columns is shown in~\figref{redundantFieldCount}.
Next, we calculate the amount of unused column data.
To do so,
we estimate columns with unbounded types 
(e.g., ``varchar'') using 2450 bytes,\footnote{\small{\cite{blog:length} 
and~\cite{comment:length} show that the most 
popular length 
of a blog is around 2450 words and the length of a comment is around 200 words. 
}} 
and we count the actual size
stored in the database for other column types (e.g., 4 bytes for ``int''). 
The results are shown in~\figref{redundantField}.

\begin{figure}[h]
  \centering
  \includegraphics[width=1\columnwidth]{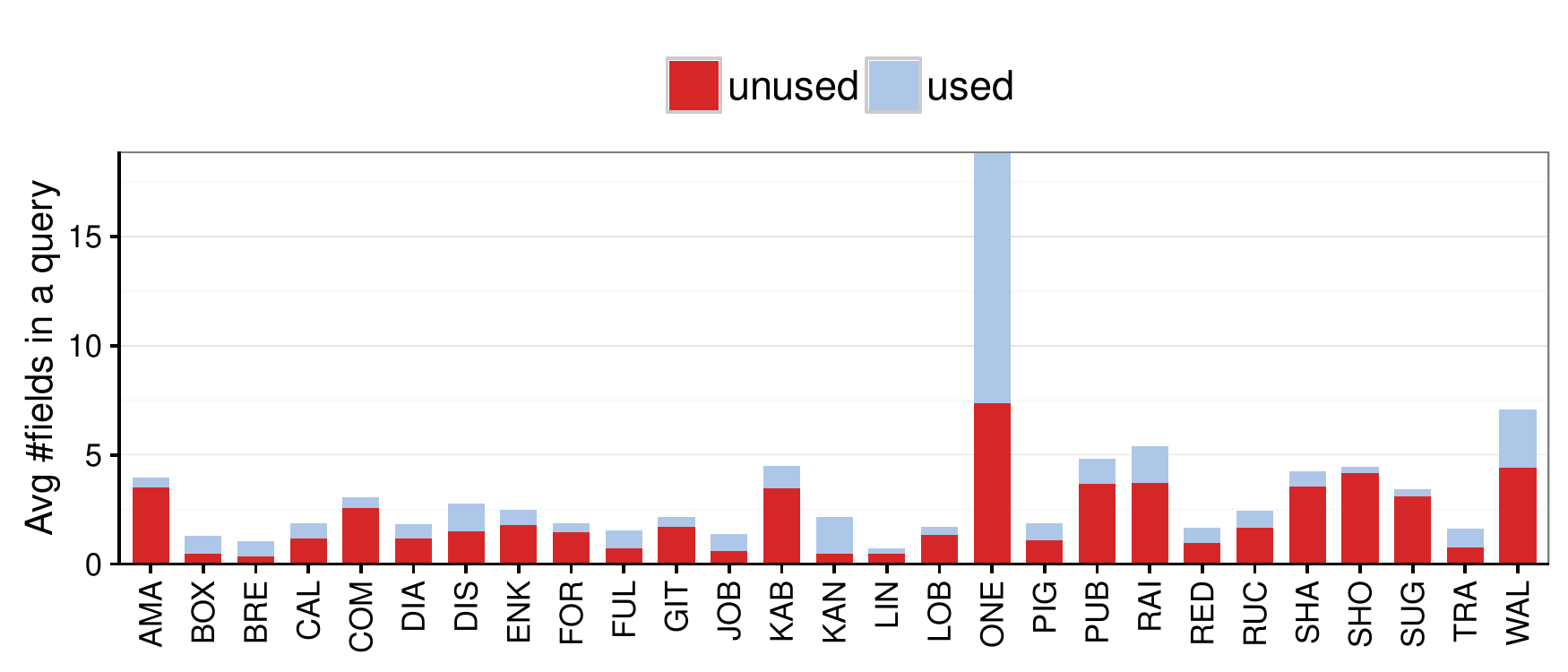}
  \caption{Number of columns retrieved that are used subsequently in the application.}
  \figlabel{redundantFieldCount}
\vspace{-0.15in}
\end{figure}

\begin{figure}[h]
  \centering
  \includegraphics[width=1\columnwidth]{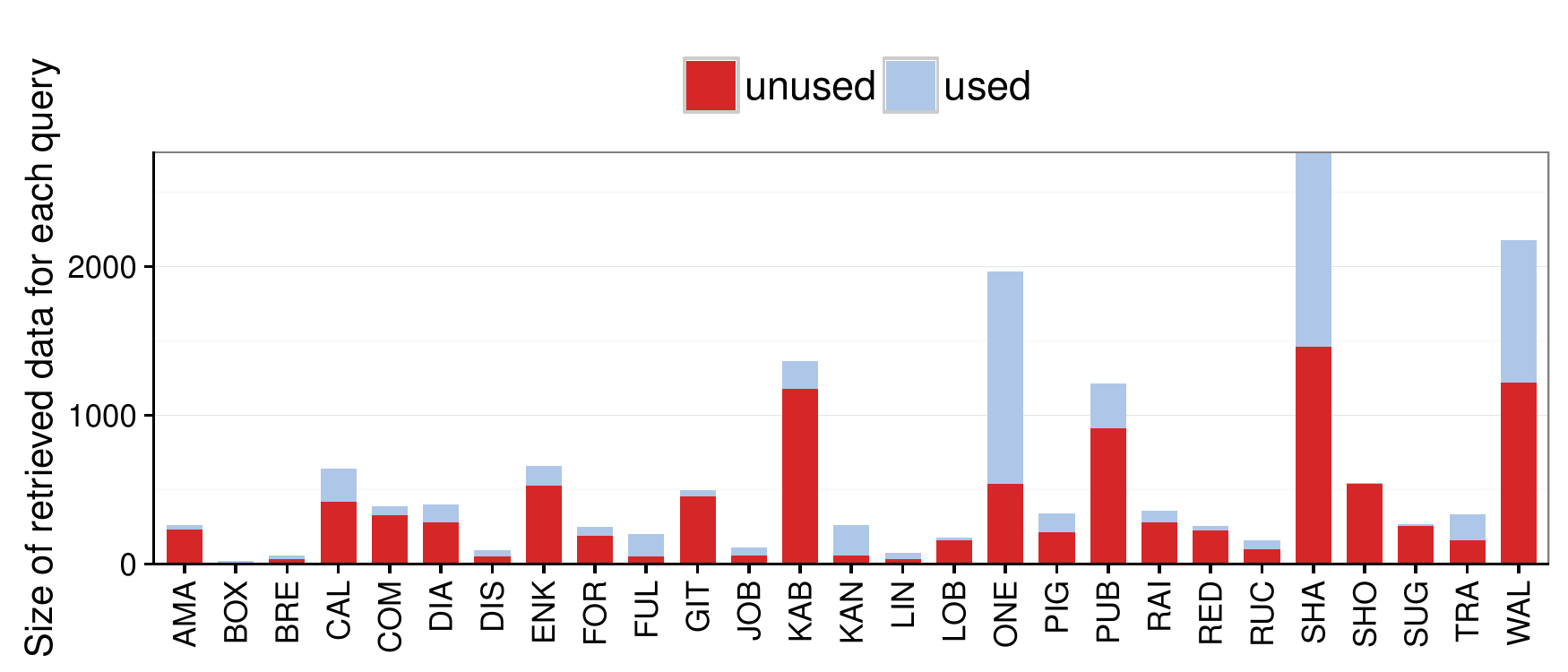}
  \caption{Amount of unnecessary data retrieved but not used subsequently 
           in the application.}
  \figlabel{redundantField}
\vspace{-0.1in}
\end{figure}

\noindent{\bf Findings \& Implications.}
We observed that
more than 63.4\% columns retrieved in applications are unused,
and these columns accounts for 62.7\% of all retrieved data as measured in bytes.

This finding indicates that
we can use static program analysis to automatically identify 
unnecessary column retrievals. 
After identifying such columns, a 
code-transformation tool can automatically change the code
so that only used columns are retrieved.
Since many ORM frameworks (including Rails) provide APIs to project certain columns
instead of all of them, such changes can be made entirely 
at application program level. 
%

\noindent{\bf Previous work.}
Prior work has studied the unnecessary column retrieval 
problem~\cite{chen:se14:antipattern} but
has {\it not} had an effective way to automatically identify them, which
\tool is capable of. In particular, it identifies such columns through a 
combination of static analysis and
dynamic profiling: static analysis analyzes ORM programs to identify 
all the used columns and dynamic 
profiling collects query logs to identify which columns are retrieved. 
However, as shown by \tool, using static analysis is already
sufficient for detecting both retrieved and used columns.
A future work is to extend \tool to not only automatically
identify but also eliminate unnecessary column retrievals through
code transformation.
 
\subsection{Unnecessary loading of object fields}
\label{sec:unne_join}

\noindent{\bf Motivation.}
Most ORMs store object fields that are nested in another object 
by normalizing the nested fields into individual tables. Join queries
are used to re-create the parent object. Similar to the previous 
section, we have also used \tool to identify unnecessary loading of 
object fields that are not used in subsequent computation. 

The source of such unnecessary loading is often due to the design
of the ORM API. For instance, all nested object fields are loaded lazily in 
Rails unless the user invokes the
{\tt includes} function.
However, since lazy loading might cause the ``N+1'' problem~\cite{Nplus1}, 
Rails encourages developers to use functions like {\tt includes} to specify 
which objects should be loaded eagerly if such information can be determined.
An example is shown in~\listref{unusedJoin} where, when a {\tt todo}
object is loaded, a query shown in~\listref{unusedJoinQuery} is issued 
to eagerly load all associated {\tt projects},
{\tt tags}, and {\tt predecessor} objects.
Among the other objects that are eagerly loaded with
the {\tt todo} object, only the {\tt predecessor} object is used in subsequent
code as shown on~\lstref{useTodo}, leaving the data retrieved from
other tables unused.

\begin{lstlisting}[language=Ruby,caption={Example of unnecessary join with lazy loading. 
Code snippet from tracks\cite{tracks}.},label={unusedJoin}]
def load_todo(cd)
  t=Todo.where(cd).includes(:projs,:tags,:preds)
  return t @\lstlabel{loadTodo}@
end
def show_recent_todos
  todos = load_todo('created > Time.now-10.days')
  pred = todos.preds  @\lstlabel{useTodo}@
end
\end{lstlisting}

\begin{lstlisting}[language=SQL,caption={Query issued by code in Listing 4},label={unusedJoinQuery}]
SELECT * from todos AS t1 WHERE created > '2016-11-01' INNER JOIN projs AS p ON p.id = t1.proj_id INNER JOIN tags AS t ON t.id = t1.tag_id INNER JOIN todos AS t2 ON t1.pred_id = t2.id
\end{lstlisting}

\noindent{\bf \tool Analysis.}
We quanitifed the number of unnecessary object loading using \tool.
First, we identified the ORM functions that explicitly issue
{\tt JOIN} queries (e.g., {\tt includes}), and analyzed how many joins
are involved in the query (three joins in~\listref{unusedJoin}).
Then we traced the dataflow graph to identify whether the joined fields are 
used (e.g., {\tt todos.predecessors} as shown in~\listref{unusedJoinQuery}), 
and counted the number of useful join if 
the associated data is used (e.g., the join in~\listref{unusedJoin}).
The results are shown in~\figref{redundantTable}.

\begin{figure}[h]
  \centering
  \includegraphics[width=1\columnwidth]{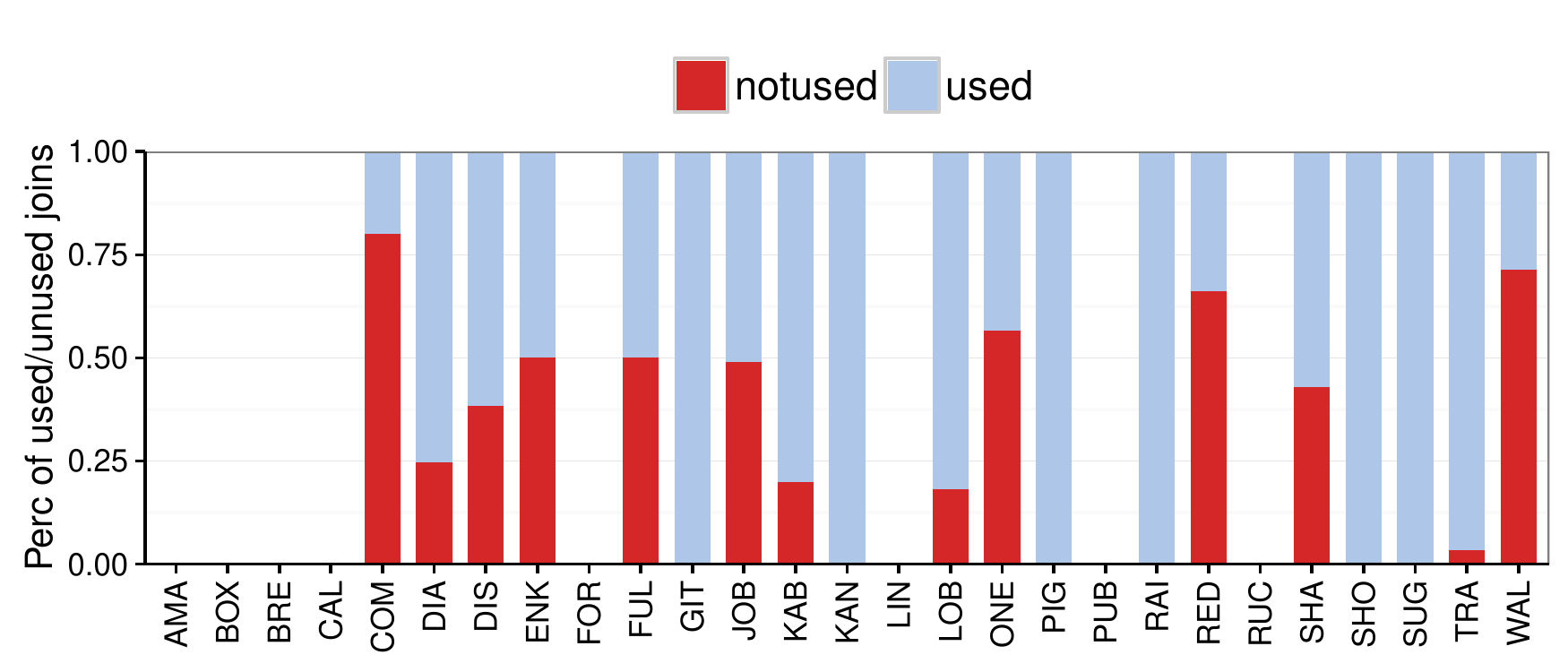}
  \caption{Percentage of unnecessary objects that are loaded eagerly across 
           different applications. No bar means that the application uses only
           lazy loading.}
  \figlabel{redundantTable}
\vspace{-0.1in}
\end{figure}

\noindent{\bf Findings \& Implications.}
Overall, we found 
32.2\% of the eager joins loading data that is not used subsequently
in the program.
We believe static program analysis similar to ours can be used to 
automatically identify whether nested objects should be loaded or not,
and eliminate such unnecessary loads.
Since many ORMs that support lazy loading by default provide API function
to specify objects to be eager-loaded,
static analysis can help determine which such calls should be used to eagerly
load the nested objects.

%

\noindent{\bf Previous work.}
Prior work~\cite{chen:finding} 
has observed similar problem in ORMs that use eager loading by default
(e.g., Hibernate~\cite{hibernate}), but {\it not} in those where lazy loading
is the default (e.g., Rails).
Rather than setting eager or lazy as the default for all classes, a better design is to
specify eager or lazy loading for each object retrieval, and 
we show that this can be done automatically. 
%

\subsection{Determining the domains of columns}
\seclabel{sec:op_constcolumn}
\noindent{\bf Motivation.}
Selectivity estimation is important for query optimization, but also
challenging to accomplish. For example,
Postgresql conducts selectivity estimation based on per-column histograms 
implemented by developers, which introduces extra programming burden.
MySQL does not conduct selectivity estimation when no index is created on
a column, and determines its query plan based on the table size, which 
could be sub-optimal. MySQL provides an easy-to-implement options to
turn on statistic collection by turning on ``AUTO\_CREATE\_STATISTICS'',
but doing so MySQL will calculate statistics for all columns, which 
gives a lot of burdens to database.

By analyzing the applications, we discover that
for many columns, developers only store constant strings or
numbers in them, and those columns are used frequently in queries.
Using program analysis can determine the domain of such columns,
and the values in the domain are usually small. 
These columns may worth creating histogram to collect statistics
since the overhead of maintaining histogram is trivial. 


Before explaining the exact program analysis algorithm and our detailed 
finding, we first look at an example 
of how developers store constants into columns.
In~\listref{todoAssign}, the {\tt state}
of {\tt todos} can only have one of the three values 
"deferred", "complete", "active"
instead of being any random string.
Similarly for the {\tt state} columns in {\tt projects} table
and {\tt contexts} table, where projcts and contexts can only have 
three states, and the domain of the {\tt state} columns includes
limited values. Maintaining histogram for those columns
incur trivial overhead since only three bins are needed for the histogram.
Knowing the statistics from histogram will enable the
database to make better query plans for queries like~\listref{todoQuery}.
 



\begin{lstlisting}[language=Ruby, caption={Assignments to state column. Code snippet abridged from tracks\cite{tracks}.}, label={todoAssign}]
class TodosController
  def create
    todo.state = "active"
  def mark_done
    todo.state = "complete"
  def mark_deferred
    todo.state = "deferred"
\end{lstlisting}

\begin{lstlisting}[language=SQL, caption={Join query on todos, contexts and projects, which may be accelerated when using histogram. Query abridged from tracks.}, label={todoQuery}]
SELECT todos.* FROM todos INNER JOIN contexts AS c ON c.id = todos.context_id INNER JOIN projects AS p ON p.id = todos.project_id WHERE todos.state = 'active' AND (NOT(c.state = 'close')) AND (p.state = 'hidden') ORDER BY todos.due
\end{lstlisting}

\noindent{\bf \tool Analysis.}
We analyze the data source for each assignment of class fields that are stored in 
table columns.
To do so, we examine each node with such assignment in the AFG in each controller action, 
and trace backwards along dataflow edges till we find a node with query function, 
or a node with no incoming dataflow edge. We refer to those as the source nodes 
and categorize them as follows: 
(1) user inputs, (2) read query functions, (3) constant values, 
(4) utility function calls (for example, Time.now),and (5) global variables.
In this section, we count the number of columns whose data sources are only constant values.
The result is shown in~\figref{fieldSource}.

\begin{figure}[h]
  \centering
  \includegraphics[width=1\columnwidth]{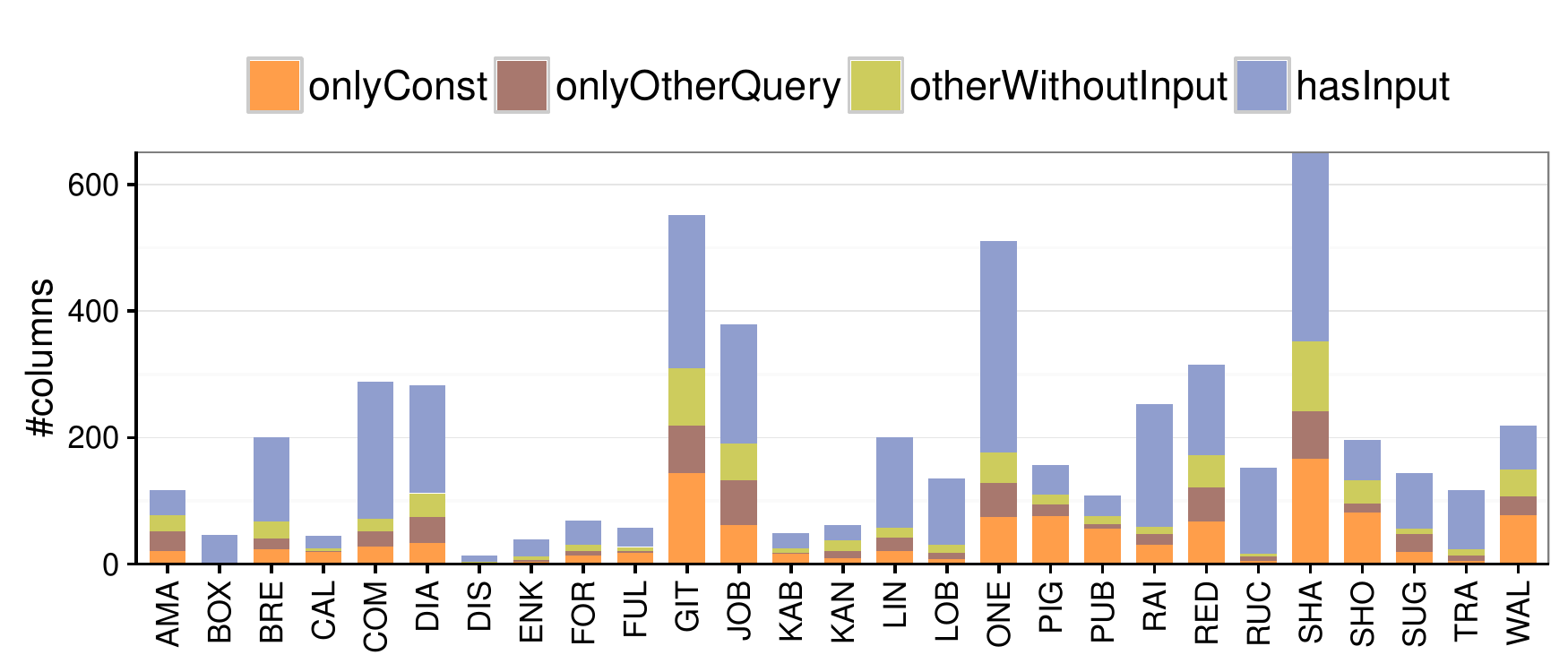}
  \caption{Breakdown of columns by their data source type. Bar 'onlyConst' represents
columns whose data sources are only constant values. Bar 'onlyOtherQuery' represents
columns whose sources are only other queries or other columns in the database.
Bar 'hasInput' represents columns that stores data computed from user inputs, and
bar 'otherWithoutInput' represents all the rest of the columns.}
  \figlabel{fieldSource}
\vspace{-0.1in}
\end{figure}

\noindent{\bf Findings \& Implications.}
The domain of 20.1\% columns can be determined from static analysis
because they stores constant values.

Program analysis can identify such columns and their domains,
and couple with existing techniques that are based on histograms.
For instance, tools can be built to automatically create histograms
for these columns. This is a better solution than maintaining no
statistics at all, or doing so for all columns, since columns of
large domain (e.g., all possible numeric values) will cause great
overhead of maintaining histograms.


\section{Physical Design}
\seclabel{physical}

In this section, we show the code patterns that reveal
opportunities for better physical design.
We have two major findings:
\begin{itemize}
\item Many applications reorganize the data retrieved from
database into other data structures like tree. 
Automatically maintaining other data structures from
relational storage can help reduce programming effort and
improve performance.
\item Values of some database columns are derived entirely from 
query results, indicating that program analysis can automatically 
 detect some functional dependencies. 
\end{itemize}

\subsection{Physical layout}
\noindent{\bf Motivations}
ORM applications store persistent data in relational database.
However, some data is not always used in a relational way.
Under these circumstances, an ORM application may need to repeatedly
retrieve data from relational database and then reorganize the data
into other data structures like tree or graph, which may greatly 
hurt performance.

\noindent{\bf Manual Study}
We manually checked the source code of all the studied ORM applications 
and found many cases of re-organizing relational data,
where tuples retrieved are put into array, and later used to build other
data structures. These data structures and applications
are listed in~\tableref{otherDataStructure}.

An example is shown in~\listref{buildTree}. 
By default, the comments retrieved from {\tt Comment} table are stored in an
array {\tt parents} (~\lstref{parent}), which is later iterated over 
and reorganized into a tree of comments
(from~\lstref{treeStart} to~\lstref{treeEnd}), with the parent-child
relationship in the tree representing a reply-to relationship. 
Such data transformation may take long time when the retrieved relational
data has a big size.
For example, the operation in~\listref{buildTree}
takes 11.32 seconds to finish for 100K returned records.

\begin{lstlisting}[language=Ruby, caption={Example of building 
a tree using returned query results. Code snippet from lobsters\cite{lobsters}. 
For 100K records, the tree-building process takes 11.32 seconds.}, label={buildTree}]
threads = Comment.where('user_id = 1').group_by(:thread_id) 
parents = Comment.where('thread_id IN #{threads}').group(:parent_comment_id) @\lstlabel{parent}@
results = []
ancestors = []
subtree = parents[nil]
#Building a tree
while subtree     @\lstlabel{treeStart}@
	if (node = subtree.shift)
		children = parents[node.id]
		node.indent_level = ancestors.length
		next unless children
		ancestors << subtree
		results << node
    subtree = children
	else
		subtree = ancestors.pop
	end
end								@\lstlabel{treeEnd}@
#Nodes are sorted in width-first traverse order
return results
\end{lstlisting}

Our manual investigation shows that the above data-structure reorganization
is demanded by program semantics.
For example, forum applications often need to organize forum comments
into a comment tree like the one shown in \listref{buildTree}, 
where a child comment is a reply to its parent comment;
project-management
applications like {\tt gitlab}\cite{gitlab} and {\tt redmine}\cite{redmine}
 often need to use trees to 
demonstrate dependencies between activities or
projects; map-related applications like {\tt openstreetmap}\cite{openstreetmap} often needs to
re-organize map tiles, which are stored in relational database, into a grid
sorted by longitude and latitude positions.

We have also observed that such data-structure re-organization could be
very inefficient and repeatedly conducted, causing huge performance loss.
For example, 
the same tree or grid or others
may need to repeatedly reconstructed at every invocation
of a controller action.
And each data reorganization could take tens of seconds or more like the
example shown in \listref{buildTree}.

\begin{table}[]
\centering
\caption{Turning query results from arrays to other data structures.}
\label{otherDataStructure}
\begin{tabular}{l|l}
Tree          & gitlab,	redmine,	lobsters,	browsercms \\
\hline
Grid          & openstreetmap,	calagator, iNaturalist        \\
\hline
Graph         & jobsworth, huginn                      \\
\hline
Interval tree & openproject                 \\
\hline
Hash map      & huginn
\end{tabular}
\vspace{-0.1in}
\end{table}

\cong{Any other prior work directly related to this?}

\noindent{\bf Findings \& Implications}
We have observed that 
many applications need to reorganize data read from relational table, 
transforming them into complicated data structures like trees, graphs, etc.

An automatic way of maintaining non-relational data structure 
atop relational storage, or physically storing data in a non-relational way,
 can greatly improve the performance of these ORM applications. 
To build such an automatic tool, one may leverage recently proposed
program synthesis techniques that can 
synthesize efficient implementations of complicated collection
data structures from high-level specifications~\cite{cozy}.
Of course, previous synthesis techniques \cite{cozy} consider
a hash or array list to be the persistent store of the original data.
In the ORM context, we will need to extend existing techniques to consider
a relational table as the persistent storage, and/or to
translate table queries into operations on hash or array list.

\subsection{Columns computed from queries}
\noindent{\bf Motivations}
Functional dependencies among columns are critical to database design and
query optimization. Traditionally functional dependency is detected via collecting 
statistics of data stored in a table.\shan{Cong, is my rephrasing about collected
statistics correct?}
For tables created by ORM applications, static program analysis is able to provide
an alternative way to detect functional dependency.

~\listref{funcdep} shows an example, where the child projects have the
same status as its parent. This code snippet reveals a conditional
functional dependency that can be detected by program analysis:
for the assignment of {\tt p.status}, the data source of that assignment
comes from a query on the column {\tt status}, if the
condition {\tt id=p.parent\_id} satisfies.
Such dependency information can be used for choosing query plan 
for queries like~\listref{funcdepSQL}.

\begin{lstlisting}[language=Ruby,caption={Example of functional dependency
that can be detected by program analysis. Code snippet abridged from redmine\cite{redmine}.},
label={funcdep}]
p = Project.new
p.parent_id = params[:parent_id]
p.status = Project.where(:id=p.parent_id).select(:status)
p.save
\end{lstlisting}

\begin{lstlisting}[language=SQL,caption={Queries that may use functional dependency
to determine query plan},label={funcdepSQL}]
SELECT * from projects AS p INNER JOIN projects AS pc WHERE p.id IN (...) AND p.status <> 1 AND p.id = pc.parent_id AND pc.status <> 1
\end{lstlisting}

\noindent{\bf \tool Analysis}
We use \tool to analyze the data source for every assignment of a class field 
that is physically stored as a table column,
as discussed in Section \ref{sec:op_constcolumn}.
We consider a column {\tt c} to have functional dependency on other columns, 
when all its data sources are query results.
That is, {\tt c} potentially has dependency on the columns involved in 
the source queries. We count the number of such columns, and the 
result is presented in~\figref{fieldSource}.

\noindent{\bf Findings \& Implications}
We have observed that
11.7\% of the columns are derived entirely from query results.
These columns are likely to be
functionally dependent on other columns.

Static program analysis could provide an alternative approach to computing
functional dependencies. Specifically,
if a column {\tt c} is computed from the results of a set of queries $\mathbb{Q}$,
further analysis on $\mathbb{Q}$ can determine
which columns $\mathbb{C}$ are involved in $\mathbb{Q}$ and 
further program dependency analysis can infer
what is the exact relationship between $\mathbb{C}$
and {\tt c}. Furthermore, if a columns is computed in different ways 
under different user inputs or user actions, the dependency relationship would
need to contain some predicates conditioning on user inputs.

\section{Query-result processing}
\seclabel{scalable}

\noindent{\bf Motivations}
How well an application scales with big data is crucial to the success
of the application. Users will probably run away from an application that
slows down dramatically when the data(base) size becomes bigger or if the
web application's response time goes beyond tens of seconds.
Since query scalability has already been studied
\cite{armbrust:sigmod13:scale}, we will study computation scalability
on ORM-application servers (i.e., excluding query time) through
a combination of static program analysis
and dynamic profiling.


\noindent{\bf \tool Profiling}
Intuitively, when database size increases, some queries would 
return larger numbers of records, causing query-result processing to
take longer time. 

To quantitatively examine such a trend of scalability or lack-of-scalability, 
we profiled seven representative ORM applications.
For each application, we record the application-server time (i.e., excluding
database query time) of different actions. 
Some actions are quick and some are slow, so for each action we plot
the server time and the number of tuples returned by queries in that action
to examine the relationship between them.
Since we do not mean to compare the execution time and query-result size
across different actions or different applications, we normalize the 
results and only show the 
{\it relative} execution time and query-result size in the figure, 
with the maximum value as baseline 1.

\noindent{\bf Profiling Findings \& Implications}
The results 
in~\figref{resultsizeTime} confirm the intuition that when an application
action retrieves larger query results, the application processing time will
increase.
There are a few exceptions, demonstrated by points on the right bottom 
corner of~\figref{resultsizeTime}.
Our manual investigation shows that most of these happen when the
query results are only used as inputs to other queries and are not 
one-by-one processed
by program (discussed in~\secref{unrequired_query}). 

Our profiling also shows that these ORM applications all suffer from
severe performance problems when the database size is big.
For example, we have measured the longest processing time (baseline 
for y-axis)
in~\figref{resultsizeTime}) in each application:
36 seconds in {\tt gitlab}, 8 seconds in {\tt kandan}, 
11 seconds in {\tt lobsters}, 311 seconds in {\tt redmine}, 
22 seconds in {\tt sugar}, and 111 seconds in {\tt tracks}, when the
action returns around 100K database records (baseline for x-axis). 
Such a slow performance,
over one minute for several applications, is intolerable for end users.
These findings indicate the importance of limiting the number of 
records returned by queries to achieve good 
application scalability.

\begin{figure}[h]
  \centering
  \includegraphics[width=1\columnwidth]{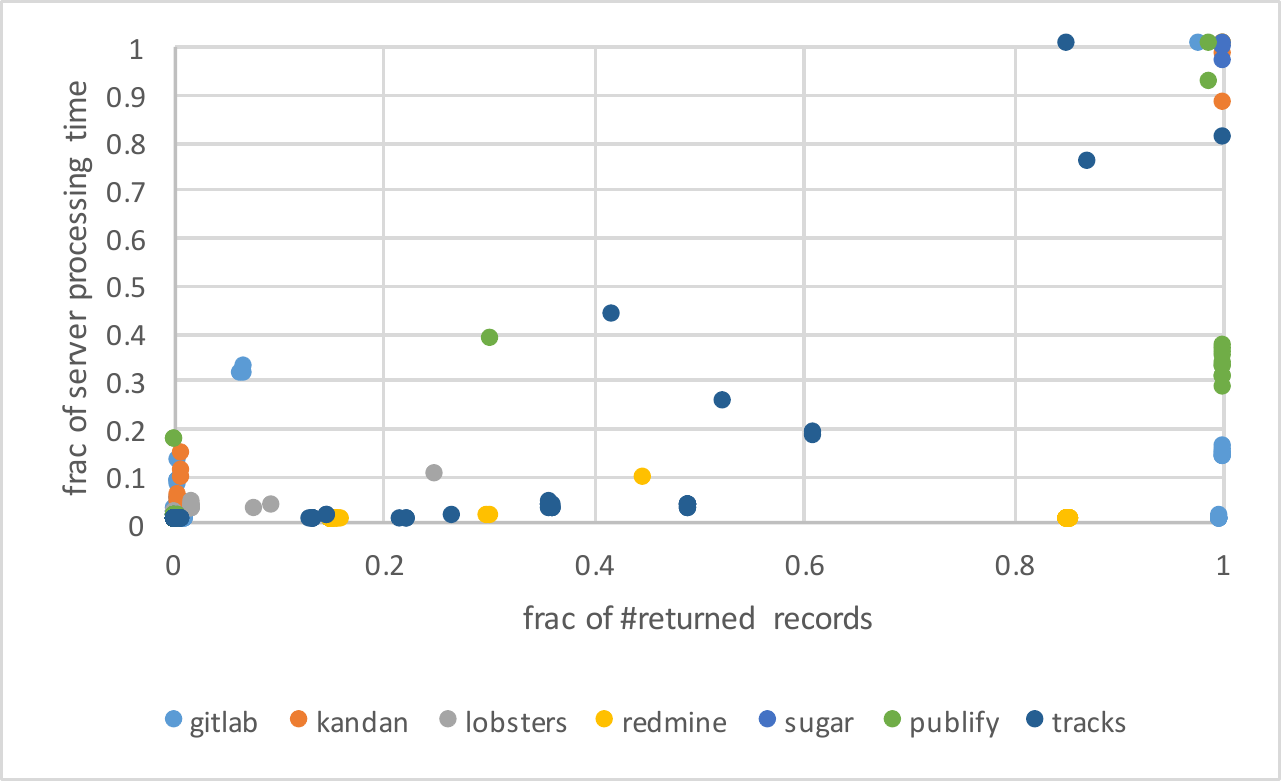}
  \caption{How server computation time increases with the number of records returned. 
Each point represents a single action, with colors differentiate applications.
The x-axis and y-axis values are normalized based on the
largest numbers of return records and the longest computation time.
}
\figlabel{resultsizeTime}
\vspace{-0.1in}
\end{figure}

\noindent {\bf \tool Analysis}
Following the above profiling finding, we then use static program
analysis to check how often ORM applications contain queries whose
result size increases with the database size.

Specifically, in Rails, a query would return unbounded-sized results
(i.e., result size increasing with the database size) in all but
the following situations: (1)
the query always returns a single value 
(e.g., a COUNT query); (2) the query always returns
a single record (e.g., retrieving using a unique identifier); (3) the query 
uses a LIMIT keyword bounding the number of returned records.
\tool static analysis goes through all queries in an ORM application
and decides whether a query has bounded or unbounded result size based on the
query type discussed above. We then
count the average number of queries returning bounded or unbounded
numbers of records, with the result shown in~\figref{queryCard}.

\begin{figure}[h]
  \centering
  \includegraphics[width=1\columnwidth]{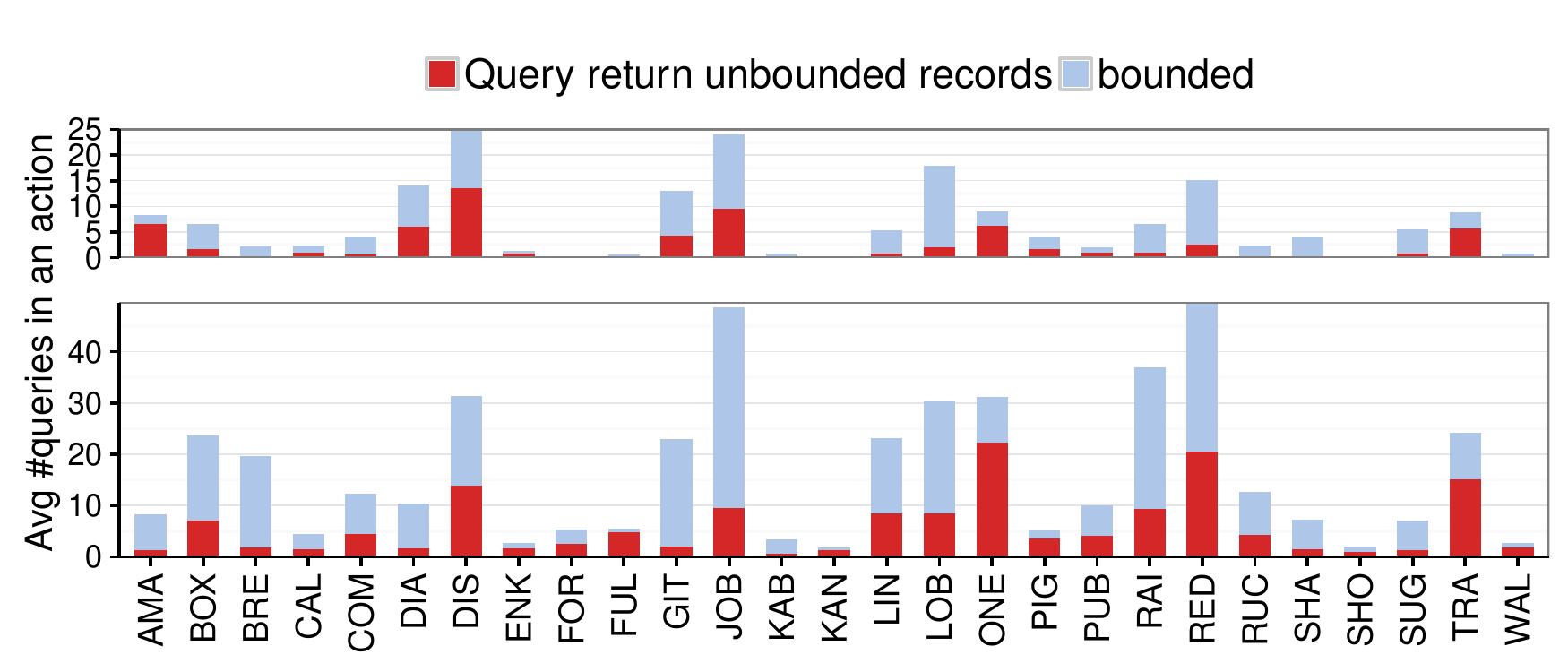}
  \caption{Queries returning bounded/unbounded \# of records}
  \figlabel{queryCard}
\vspace{-0.15in}
\end{figure}

\noindent {\bf Analysis Findings \& Implications}
We found that
many queries (35.5\%) return unbounded numbers of result
records.
Processing these queries is likely to become scaling bottleneck.

Turning queries from returning unbounded results to bounded can clearly help
improve application scalability. However, it would require changes
to application code. 
For instance, if an application
designer wishes to show all the comments on a single webpage, 
she will write code to retrieve all comments and render them, 
making the response time of the corresponding webpage to increase at least 
linearly with the {\tt comments} table size.
An alternative design is to render only $K$ comments at a time,
and incrementally load more comments when the user scrolls down the page.
In this case the query on the {\tt comments} table returns a small number of
records.
We manually tried the above changes for actions in two applications,
{\tt redmine} and {\tt publify}. 
These changes lead to huge performance improvement:
98\% reduction in initial response time for {\tt redmine} and 50\% for 
{\tt publify} as shown in~\figref{pageScroll}.

When we change an application to use incremental loading, two major challenges
remain. First, queries needs to be rewritten to retrieve only the data shown 
initially on the webpage. 
What data to present on the initial page and what to load next
depend on user interaction. For instance,
the loading method for scrolling (initially showing top tuples) 
differs from zooming (initially showing aggregate results).
Second, queries need to be further optimized to reduce the overhead
of each loading (e.g., loading in each page scroll). 
For instance, in {\tt publify} 
the articles to be sorted by their create time before being retrieved and rendered,
which is time consuming for every scrolling.
Similar optimizations as mentioned in~\secref{sec:partial_pred} 
can be applied to accelerate queries in incremental loadings.


Overall, scalability bottlenecks caused by unbounded query results widely
exist in ORM applications. They are a severe threat to the usability
of ORM applications when data size increases.
How to redesign applications and optimize 
query execution to eliminate these scalability bottlenecks is critical
and also challenging.

\begin{figure}[h]
  \centering
  \includegraphics[page=1, width=1\columnwidth]{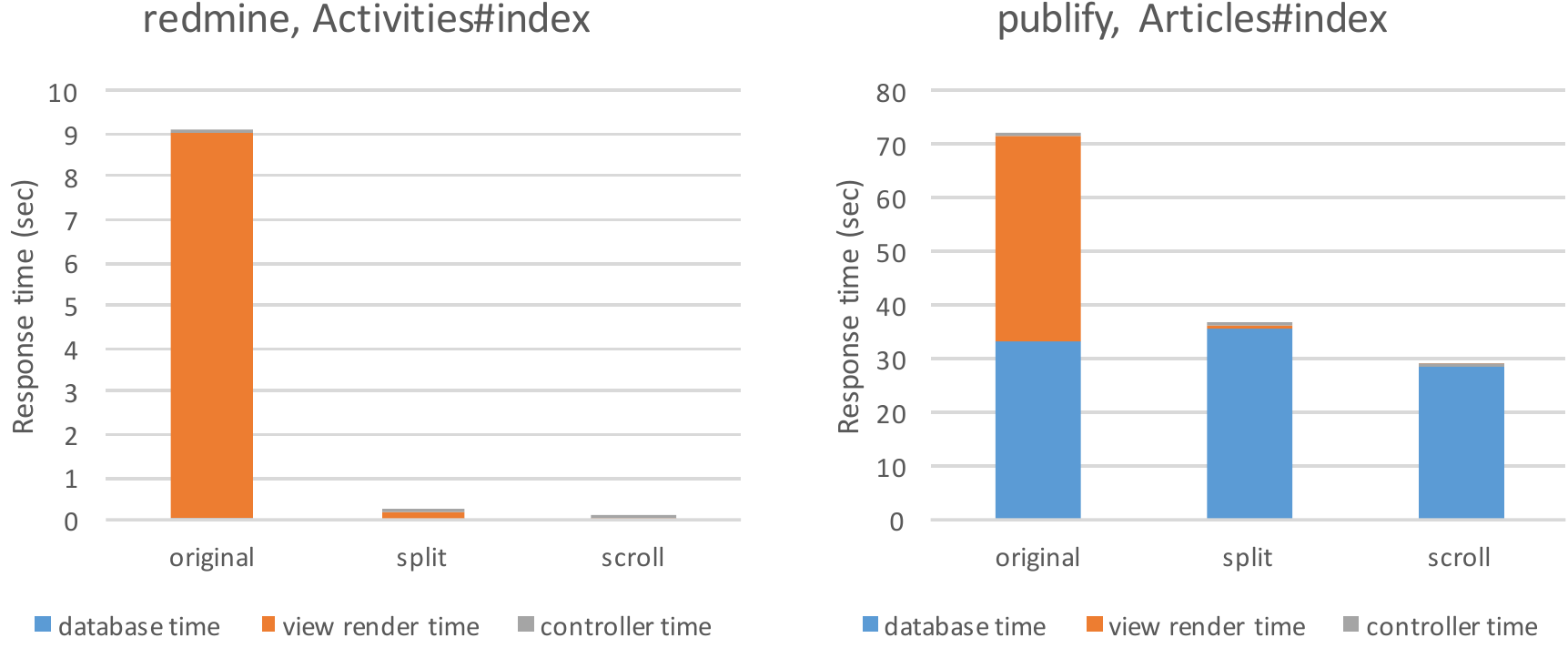}
  \caption{Performance improvement of incremental data loading on 
Activities.index action from {\tt redmine}\cite{redmine} (20K records rendered on a single page) and Articles.index action from {\tt publify}\cite{publify} (80K records rendered). 
Each subfigure shows from left to right the response time of 
the original page without incremental
loading,  the page shows 40 records under incremental loading, 
each scroll that incrementally appends
40 records to the page. }
  \figlabel{pageScroll}
\vspace{-0.1in}
\end{figure}

\section{Caching and Prefetching}
\seclabel{caching_and_prefetching}

This section studies the opportunities for 
query caching and prefetching.
Caching stores the results of previously executed queries,
which may serve later queries and avoid redundant
database execution. Since
we already discussed caching intermediate results within an action
in~\secref{sec:partial_pred}, our following study will be about inter-action
caching ---
how queries from previous actions can help queries in later actions.

Prefetching pre-executes queries which are predicted to be executed
in subsequent actions.
That is, if one can predict which web page a user will visit next, 
some queries that would be issued by the next page could be pre-executed
to reduce the response time of the next page.

We study the optimistic performance gain of caching and prefetching
in one user session, and the chances to cache and prefetch specific
types of queries.
Our major findings include:
\begin{asparaitem}
\item Caching can help serve a big portion of read queries,
but its potential performance gain is not as much, as many
queries that can benefit from caching take little time to execute.
\item Most queries (more than 90\%) issued by a web-page
can be 
prefetched as their inputs are known at the previous page.
\item Many queries issued by a page share same templates with 
some queries issued by previous pages. 
These queries can be efficiently prefetched
by combining them with queries from previous pages.
\end{asparaitem}

\subsection{Caching}
\seclabel{caching}

\noindent {\bf Motivations}
By default Rails clear the query cache after an action finishes, 
so no data is shared across actions.
However, query results of a current action can be reused to serve queries in later actions.
Furthermore, we have observed two patterns in Rails applications that 
facilitate many syntactically-equivalent queries to be issued across actions.
Caching could be beneficial for these queries.
First, Rails support filters --- the code, including queries, inside a filter 
$f$ of class $C$ would be executed every time a member function of $C$ is 
invoked. Consequently, the same queries in $f$ would be issued by many different 
actions that invoke member functions of $C$.
Second, many pages share the same layout. Consequently, the same queries are
repeatedly issued to generate the same layout when rendering different pages.

We aim to better understand how beneficial inter-action caching can be
for ORM applications, which will then reveal whether it is worthwhile to
design a good inter-action caching mechanism.
We mainly focus on the caching opportunity within a user session, and leave
multi-session and multi-user analysis as future work.

\noindent {\bf \tool Profiling}
To study the optimistic performance gain by inter-action caching, we will 
simulate a user session by randomly visiting 9 pages\footnote{\small{We choose 9 
to be the number of pages per user session (not including the login page)
since the average number of visits per day per user for the 10 most popular website in
the U.S. is 8.35~\cite{alexa}.}} starting from a random page in the application, and
then compare queries issued by one action with those by previous actions from 
two perspectives.

First, we compare query results across actions. For each query, we check
whether all its result tuples can be found in the data retrieved by 
some queries from earlier actions in the same session. 
If yes, this query can benefit from caching and is categorized as 
a ``hit'' query, represented by the red bars in ~\figref{cachingQNumber} and~\figref{cachingQTime}. 

Second, we compare query strings across actions.
For each query {\tt Q}, we check whether a query with exactly the same query 
string has been issued by previous actions in this session.
If yes, {\tt Q} is considered as having a syntactically equivalent peer and
is a good candidate for using cache.
We further check whether the result of {\tt Q} is the same as its syntactically
equivalent peer {\tt Q'} or not, as the database content could have been updated 
between {\tt Q'} and {\tt Q}.
The results are represented by the two darkest blue bars in 
\figref{cachingQNumber} and~\figref{cachingQTime}.

\noindent {\bf Findings and Implications}
We show the breakdowns of the number of all queries and the execution time of all queries
in ~\figref{cachingQNumber} and~\figref{cachingQTime} respectively, following 
the two categorizing perspectives discussed above.

At the first glance, caching looks promising based on the query counts.
For all applications profiled, on average 60\% of read queries can obtain their
results from inter-action cache (i.e., hit queries); 
on average 43\% of read queries have
syntactically equivalent queries issued by earlier actions, 
among which 10\% can directly use the earlier results.

However, caching is actually not so beneficial in terms of
query running time. Only 23.6\% of query time is spent by hit queries
and only 17.6\% spent by queries with syntactically-equivalent peers.

Our findings reveal that although many queries can obtain their results from 
an ideal inter-action cache, 
these queries are usually short-running ones. 
Consequently, 
although it still depends on the type of application,
when the system bottleneck is query execution
instead of network, inter-action caching within a user session is not worthy 
to implement for most applications. 

\begin{figure}[h]
  \centering
  \includegraphics[width=1\columnwidth]{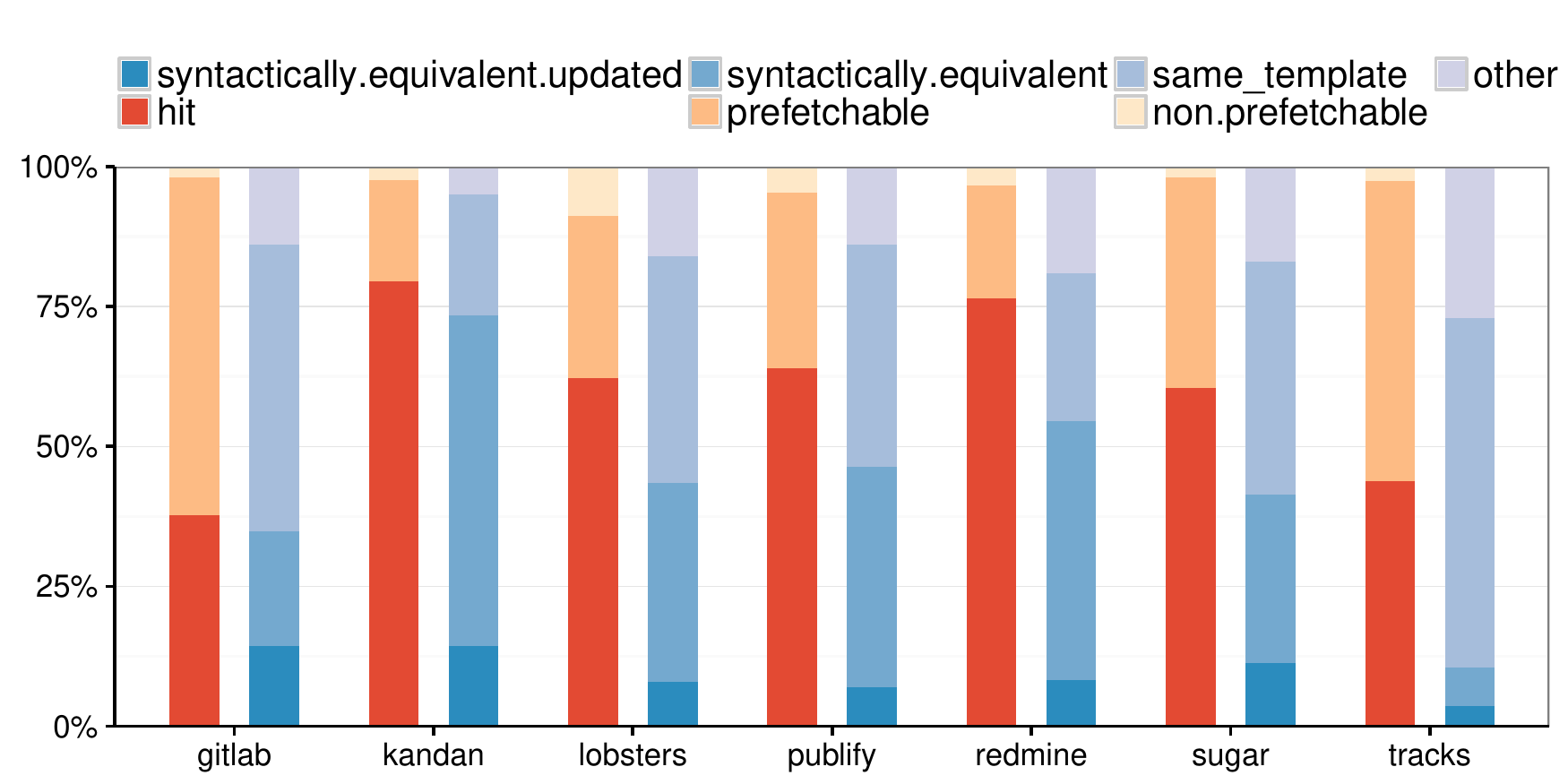}
  \caption{Two breakdowns of read queries in terms of query number. Calculated
from profiling results.}
  \figlabel{cachingQNumber}
\vspace{-0.1in}
\end{figure}

\begin{figure}[h]
  \centering
  \includegraphics[width=1\columnwidth]{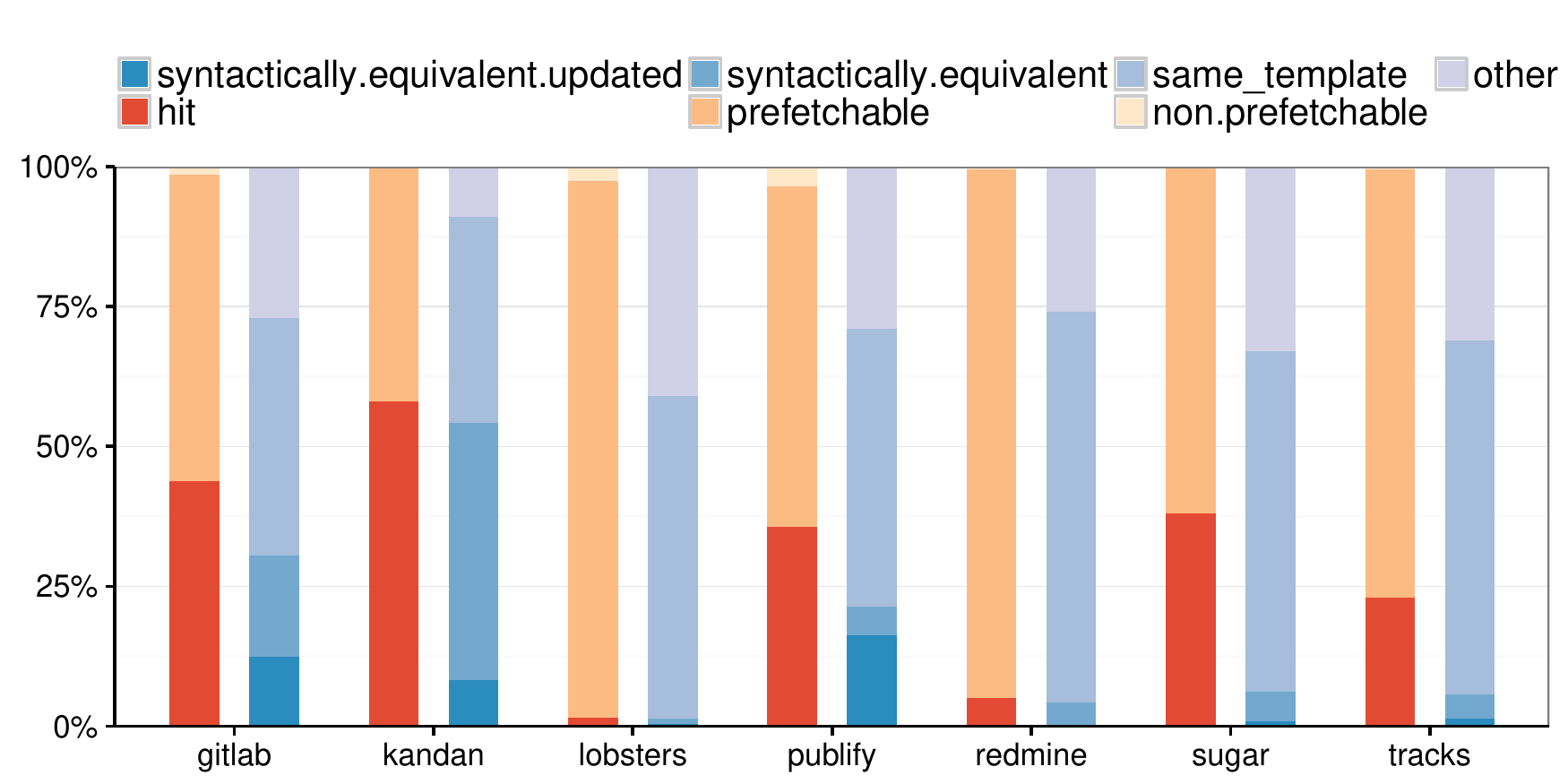}
  \caption{Two breakdowns of read queries in terms of query running time. Calculated
from profiling results.}
  \figlabel{cachingQTime}
\vspace{-0.1in}
\end{figure}

\subsection{Prefetching}
\noindent {\bf Motivations}
In this section we study the potential performance gain of prefetching. 
Similar to~\secref{caching}, we first analyze the optimistic performance improvement by
prefetching (optimistically how many queries can be prefetched). We consider a query
to be ``prefetchable'', if all its parameters can be determined at the previous action.
Then we look into a specific type of prefetchable queries --- prefetchable 
queries from different actions that share the same template.

An example of queries sharing the same template
 is shown in~\listref{prefetchPage}, 
where 40 sorted stories 
are listed on each page.
When a user visits the first page, the query shown  
in~\listref{prefetchPageQ1} is issued.
Then, when the user visits the second page, the query in~\listref{prefetchPageQ2} is issued.
These two queries issued by consecutive actions
share the same template but contain different parameters.
Note that, the query shown in~\listref{prefetchPageQ2} is considered 
prefetchable, because all the query parameters can be determined once we predict
that page 2 ({\tt page\_id} is 1) will be visited next. 

We pay special attention to the above type of queries, because the 
same-template property allows their prefetching to be
efficiently done by combining them with queries from earlier actions.
Such combinig brings non-trivial performance gain.
For instance,
the query in~\listref{prefetchPageQ1} and~\listref{prefetchPageQ2}
can be combined as shown in~\listref{prefetchCombine}. The combining eliminates
the redundant sorting work of 
query in~\listref{prefetchPageQ2}. Consequently, executing
the combined query is faster than executing the two separate queries
one by one.


\begin{lstlisting}[language=Ruby,caption={Visiting posts listed by page. Code snippet abridged from sugar\cite{sugar}}, label={prefetchPage}]
class PostsController
  def index
    stories = Post.order('created_at').limit(40).offset(40*param[:page_id])  @\lstlabel{storyOrder}@
  end
end
\end{lstlisting}
\vspace{-0.1in}

\begin{lstlisting}[language=SQL, caption={Queries issued by Listing 11, visiting page 1}, label={prefetchPageQ1}]
SELECT * from posts ORDER BY created_at LIMIT 40 OFFSET 0
\end{lstlisting}
\vspace{-0.1in}

\begin{lstlisting}[language=SQL, caption={Queries issued by Listing 11, visiting page 2}, label={prefetchPageQ2}]
SELECT * from posts ORDER BY created_at LIMIT 40 OFFSET 40
\end{lstlisting}
\vspace{-0.1in}

\begin{lstlisting}[language=SQL, caption={Combining queries from Listing 12 and 13}, label={prefetchCombine}]
SELECT * from posts ORDER BY created_at LIMIT 80 OFFSET 0
\end{lstlisting}

\noindent {\bf \tool Profiling and Analysis}
We first analyze what queries can be prefetched,
which depends on both the query itself and the action it resides in.
As mentioned in~\secref{rails_app},
when visiting a webpage, a user can send another request to the server by either
clicking a link on the current page, or filling and submitting a form. In the former case,
the parameters to the next action are known once we predict which link the user will click.
Consequently, the next action
can be pre-run and all queries in the next action can be prefetched. In the latter case,
the parameters to the next action come from the unpredictable content that the user 
will fill in the form. Consequently, not all queries in the next action can be prefetched
--- only those whose parameters do not depend on unpredictable user inputs are 
prefetchable. 

We then check which prefetchable query shares the same template with a
query issued by the previous action, because these queries can potentially be
more efficiently prefetched through query combining as discussed earlier.

We do the above counting through both \tool profiling and \tool static analysis.
During profiling, for an action triggered by a 
GET http request, we treat all the queries issued by it as prefetchable;
 for an action triggered by a POST request, we treat only those queries
that do not use user inputs (i.e., parameters in the POST
request) as prefetchable. 
We then analyze the query log across actions.
For each prefetchable query {\tt Q}, we go through all queries from the previous
action to see if there exists a query {\tt Q'} that shares the same query template.
The results of this counting is shown
in~\figref{cachingQNumber} and~\figref{cachingQTime}.

In static analysis,
\tool gets possible
{\it current}--{\it next} action pairs linked by the 
{\it next action} edge described in~\secref{staticAnalysis}, 
as well as whether a next action is invoked through
GET or POST request. 
Inside each action, \tool marks the query that does not use user
input as parameters (analysis of query source is described
in~\secref{sec:op_constcolumn}) as prefetchable.
For a prefetchable query {\tt Q} in a {\it next} action, we check if 
there exists a query {\tt Q'} from the corresponding {\tt current} action 
that is issued by the same piece of code as {\tt Q} 
(e.g., queries in~\listref{prefetchPageQ1} and~\listref{prefetchPageQ2} are both
issued by code on~\lstref{storyOrder} of~\listref{prefetchPage}).
If {\tt Q'} exists, {\tt Q} is considered to be a prefetchable query sharing the 
same template with a previous query.
The statistics for all applications are shown in~\figref{nextAction}. 

\begin{figure}[h]
  \centering
  \includegraphics[width=1\columnwidth]{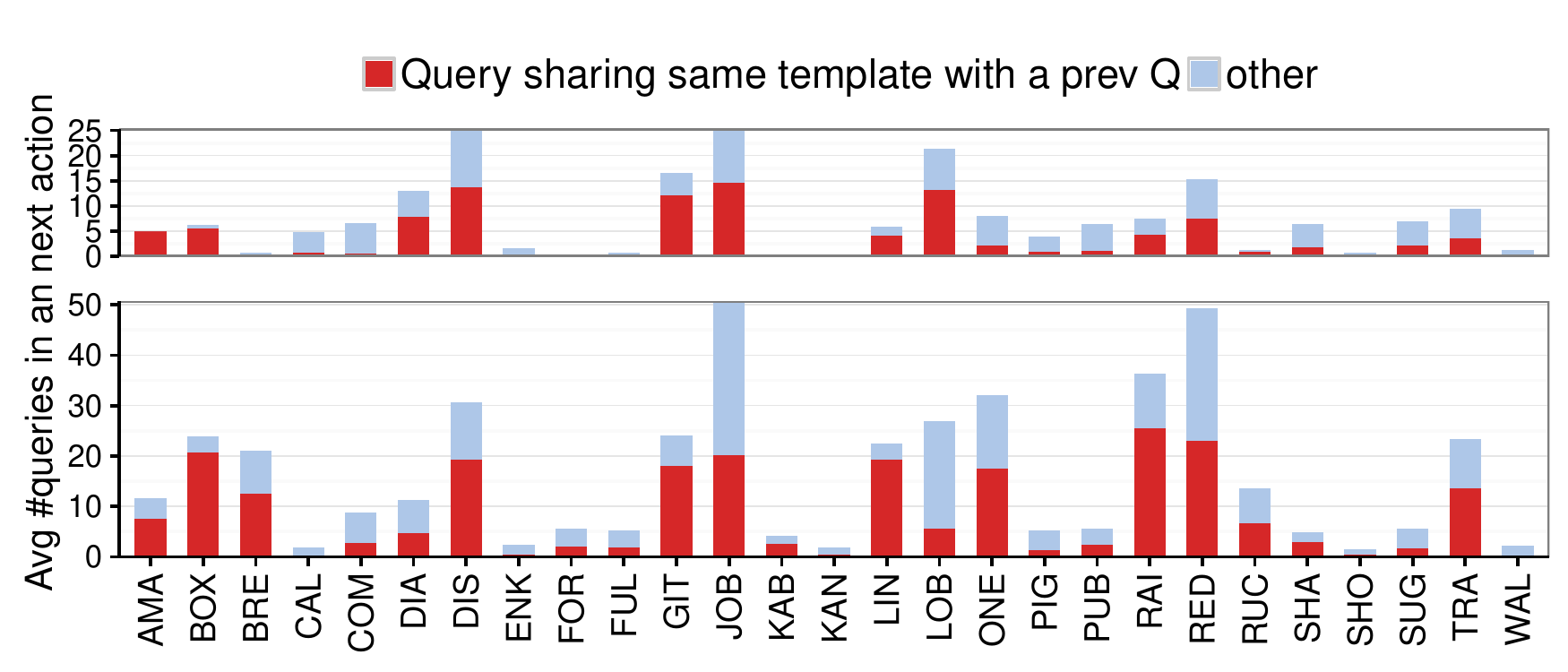}
  \caption{Prefetchable queries sharing the same template with queries in 
previous action (obtained by \tool static analysis)}
  \figlabel{nextAction}
\vspace{-0.1in}
\end{figure}


\noindent {\bf Findings \& Implications}
Intuitively, a webpage usually contains more clickable links than
forms to be filled. Consequently, most queries are prefetchable. 
This is confirmed by the 
results in~\figref{cachingQNumber} and~\figref{cachingQTime}, 
which show that most queries (93.6\% in number and 98.7\% in time) are prefetchable.
The figures also show that many queries 
(40.4\% in number and 54.3\% in time) are not only prefetchable but also
share the same template with queries from the previous action.
A similar trend is also shown by static analysis in \figref{nextAction}: 
for each pair of {\tt current-next} actions, 
53.2\% of queries issued in the next action share the same template
with a query in the current action.

The above results reveal a great opportunity of improving performance by prefetching,
and particularly by prefetching queries with the same template.
Furthermore, since static program analysis is able to identify prefetchable
queries with the same template, 
a query rewriting tool can use techniques proposed by~\cite{chavan:db11:dbridge}
to automatically combine queries to prefetch data. 
However, it can be unrealistic to prefetch queries
in all possible next actions if there are many links on a current page. 
A good prediction based on user behavior patterns will help 
only prefetch queries for actions that are most likely to be triggered next.

\section{Software testing}
\seclabel{testing}

\noindent {\bf Motivation}
Functional testing is crucial for all types of software, consuming
30\% of software development resources \cite{testingresource}.
The quality of functional testing is often measured by code coverage, such
as statement coverage and branch coverage. 
Much research has been conducted to help generate 
test inputs that can drive the testing to achieve high coverage~\cite{kleeosdi08}.
Unfortunately, existing input-generation techniques do not 
consider how to generate database content. 
Therefore, we want to examine how many branches' outcomes depend on
database content and figure out whether traditional testing techniques
are sufficient for ORM-based software.

\noindent {\bf \tool Analysis}
For every branch $b$ inside a controller action, we want to know whether the
outcome of its branch condition depends on database content or not.
To do so, we track the data flow on AFG and identify all the source 
nodes of the branch condition in $b$. If any of these source nodes includes 
database queries, the outcome of this branch could depend on database content 
and this branch is categorized as DB-sensitive. 
The result of our study is shown in~\figref{branch}. 

\begin{figure}[h]
  \centering
  \includegraphics[width=1\columnwidth]{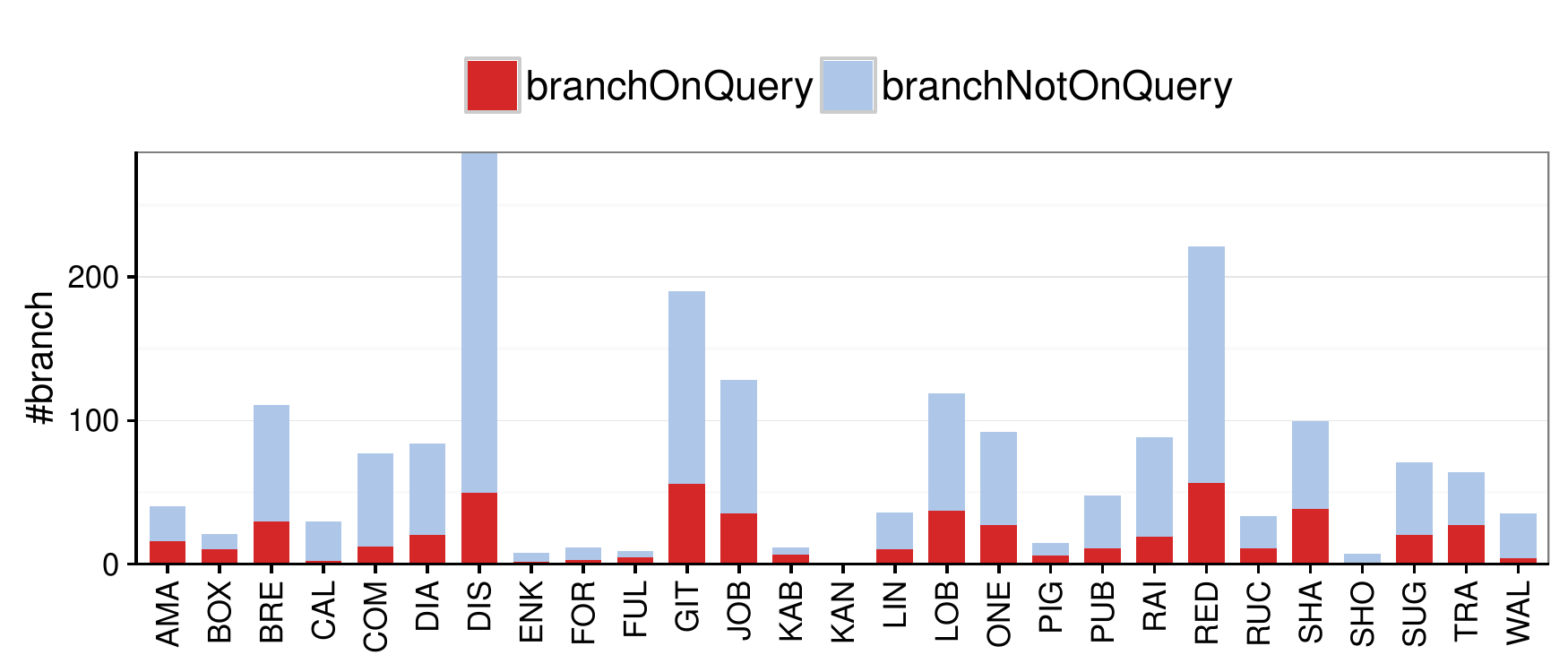}
  \caption{Average number of branches in each controller action}
  \figlabel{branch}
\vspace{-0.1in}
\end{figure}

\noindent {\bf Findings \& Implications}
We have observed that
a significant portion of branches (26.5\%) are DB-sensitive.

To achieve high testing coverage for ORM-based applications, 
one may need to carefully synthesize database content to cover both
outcomes of DB-sensitive branches.
One can potentially automated part of this by extending 
symbolic-execution based input generation \cite{kleeosdi08,dart,cute}
with query modeling.

\section{related work}
Besides the prior work related to our findings, as mentioned
in previous sections, we discuss two more major categories
of related work.

\textbf{Empirical studies}
A previous empirical study
\cite{chen:se14:antipattern} investigated performance anti-patterns for
ORM applications. However, this work only includes 
two anti-patterns, and only covers three applications using Hibernate ORM.
This paper provides a much thorough study of performance problems in
Rails-based ORM applications. 

Another previous work
\cite{bailis:sigmod15:feral} also studied open-source Rails applications like
this paper does. However, it only focuses on how Rails applications
use different concurrency control mechanisms.
This work only studied the patterns in the code,
while we performed a deeper study on these applications
using program analysis and profiling.

\textbf{Program analysis for database optimization}
Our work shares the same optimization philosophy with some 
recently proposed techniques --- optimize the 
database-related applications based on program analysis instead of query log.
DBridge~\cite{sudarshan:sql,dbridge:decorrelation,sudarshan:batch,guravannavar:vldb08:rewriting,chavan:db11:dbridge} 
includes a series of work on holistic optimization. 
Besides query batching and binding as we mentioned earlier,
this set of work also includes automatic transforming of regular object-oriented code
into synthesized queries, decorrelation of user functions and queries, etc.
Other holistic optimizations includes but not limited to,
QBS~\cite{cheung:pldi13:synthesis,sudarshan:sql} for query synthesis, 
QURO~\cite{quro} for query reordering in transactions,
PipeGen~\cite{pipegen} for automatic data pipe generation, etc.

\section{conclusion}
In this paper we studied the performance patterns on a set of 27 real-world
web applications built with ORM framework 
We build \tool to perform static analysis on these applications and examine
how the application interact with databases using ORM. \tool also generates
synthetic data and profiles the application at runtime to better understand
their performance.

Our findings reveal many optimization opportunities,
for instance,
program-analysis assisted multi-query optmization, automatic program 
transformation to reduce redundant data retrieval, determining domains/constraints 
of database columns by static analysis, automatic incremental loading,
query combining for prefetching, etc.
We point out that program analysis can play important role in many optimizations.

Future research can benefit from our study in various aspects. 
For example, web developers can use \tool to detect performance issues
and apply solutions in our study. ORM designers can leverage program analysis
to automate optimizations within ORM layer.
Database designers can gain insights on how people use databases from our study
and implement optimizations that potentially benefit a large number of applications.

\begin{small}
\bibliographystyle{abbrv}
\bibliography{ref,confs}  
\end{small}

%
%
%
%

%
%



%
%
%
\end{document}